\documentstyle[prd,aps,preprint,epsfig]{revtex}
\newfont\fiverm{cmr5}
\input prepictex
\input pictex
\input postpictex

\begin{document}

\newcommand{\TeV}{\,{\rm TeV}}
\newcommand{\GeV}{\,{\rm GeV}}
\newcommand{\MeV}{\,{\rm MeV}}
\newcommand{\keV}{\,{\rm keV}}
\newcommand{\eV}{\,{\rm eV}}
\def\ap{\approx}
\def\bea{\begin{eqnarray}}
\def\eea{\end{eqnarray}}
\def\beqar{\begin{eqnarray}}
\def\eeqar{\end{eqnarray}}
\def\ler{\lesssim}
\def\gtr{\gtrsim}
\def\beq{\begin{equation}}
\def\eeq{\end{equation}}
\def\haf{\frac{1}{2}}
\def\plb#1#2#3#4{#1, Phys. Lett. {\bf #2B} (#4) #3}
\def\plbb#1#2#3#4{#1 Phys. Lett. {\bf #2B} (#4) #3}
\def\npb#1#2#3#4{#1, Nucl. Phys. {\bf B#2} (#4) #3}
\def\prd#1#2#3#4{#1, Phys. Rev. {\bf D#2} (#4) #3}
\def\prl#1#2#3#4{#1, Phys. Rev. Lett. {\bf #2} (#4) #3}
\def\mpl#1#2#3#4{#1, Mod. Phys. Lett. {\bf A#2} (#4) #3}
\def\rep#1#2#3#4{#1, Phys. Rep. {\bf #2} (#4) #3}
\def\lpp{\lambda''}
\def\ccg{\cal G}
\def\slash#1{#1\!\!\!\!\!/}
\def\rpv{\slash{R_p}}

\setcounter{page}{1}
\draft
\preprint{KAIST-TH 01/03, KIAS-P01012}

\title{Probing the messenger of supersymmetry breaking  \\
by the muon anomalous magnetic moment}

\author{
Kiwoon Choi$^a$, Kyuwan Hwang$^a$, Sin Kyu Kang$^b$, 
Kang Young Lee$^b$, and Wan Young Song$^a$}

\address{
$^a$Department of Physics,
Korea Advanced Institute of Science and Technology\\
        Taejon 305-701, Korea\\
$^b$ School of Physics,
Korea Institute for Advanced Study, Seoul 130-012, Korea}


\tighten

\maketitle

\begin{abstract}
Motivated by the recently measured muon's
anomalous magnetic moment $a_{\mu}$, we examine
the supersymmetry contribution to $a_{\mu}$ in
various mediation models of supersymmetry breaking
which lead to predictive flavor conserving soft
parameters at high energy scale.
The studied models include dilaton/modulus-mediated models
in heterotic string/$M$ theory,
gauge-mediated model, no-scale or gaugino-mediated model,
and also the minimal and deflected anomaly-mediated models.
For each model, the range of $a^{\rm SUSY}_{\mu}$ allowed by
other experimental constraints, e.g. $b\rightarrow
s\gamma$  and the collider bounds on superparticle masses,
is obtained  together with the corresponding parameter
region of the model.
Gauge-mediated models with low messenger scale
can give any $a^{\rm SUSY}_{\mu}$ within the $2\sigma$ bound.
In many other models,
$b\to s\gamma$ favors $a^{\rm SUSY}_{\mu}$
smaller than either the $-1\sigma$ value ($26\times 10^{-10}$)
or the central value ($42\times 10^{-10}$).

\end{abstract}



\section{Introduction}

Weak scale supersymmetry (SUSY) is perhaps the most promising  candidate
for physics beyond the standard model (SM) \cite{nilles}.
Any realistic supersymmetric model at the weak scale
contains explicit but soft SUSY breaking terms which are
presumed to originate from some high energy dynamics.
If one writes down the most general form of soft
terms, 
it would require too many parameters, e.g. 
more than 100 even for the minimal supersymmetric standard model (MSSM).
Furthermore, for a generic form of soft terms,
the superparticle masses should exceed about 10
TeV in order to avoid dangerous flavor changing processes \cite{fcnc}.
Such large superpartner masses spoil the natural emergence
of the weak scale, and thus the major motivation for 
supersymmetry also. 

In view of these difficulties of generic soft terms,
it is quite demanding to have a theory of soft terms
leading to a predictive form of flavor
conserving  soft terms.
In fact, the shape of observable soft terms
is mainly determined by the mediation mechanism
of SUSY breaking, i.e. by the couplings of the SUSY
breaking messenger fields
to the observable fields, rather than by 
the SUSY breaking dynamics itself.
This is a good news since in many cases the couplings of the
messenger fields can be treated in perturbation theory,
while the SUSY breaking dynamics involves nonperturbative
effects.
Therefore once the messegers of SUSY breaking are identified,
one can get a well-defined prediction for soft parameters.
As long as the predicted soft parameters conserve the flavors,
their size can be of order the weak scale.
This  would allow the prediction to be tested by the future collider
experiments and/or the low energy precision experiments.
There already exist many interesting proposals for flavor conserving
soft parameters, e.g. dilaton/modulus mediation
in heterotic string/$M$ theory \cite{string,mtheory}, 
gauge mediation \cite{gauge},
no-scale \cite{noscale} or gaugino mediation \cite{gaugino},
anomaly mediation \cite{anomaly,moreanomaly},
and others \cite{others}.

Very recently, the BNL experiment E821 has reported
a measurement of the muon's anomalous magnetic moment,
indicating a $2.6\sigma$ deviation of $a_{\mu}\equiv
(g_{\mu}-2)/2$ from the standard model value \cite{muon}:
\beq
\label{deviation}
\Delta a_{\mu}\equiv a_{\mu}^{\rm exp}-a_{\mu}^{\rm SM}=
(42\pm 16)\times 10^{-10}.
\eeq
Although can be consistent with the standard
model value if one takes other theoretical calculations
of the hadronic vacuum polarization \cite{yndu}, 
this  may indeed be a sign of new physics beyond 
the standard model.
In particular, this deviation can easily find  its explanation
in supersymmetric models through the well-known
neutralino-smuon and chargino-sneutrino diagrams \cite{marciano}.
An explicit formula  of the SUSY contribution 
to  $a_{\mu}$ is presented for instance 
in Ref. \cite{moroi}.
The SUSY contribution to $a_{\mu}$ is enhanced as $\tan \beta$
increases, and the chargino-sneutrino diagram provides a dominant
contribution for generic SUSY parameters.
In the limit of degenerate superparticle masses,
the leading contribution is approximately given by
\cite{marciano}
\begin{equation}
a^{\rm SUSY}_{\mu} \approx \frac{\alpha(M_Z)}{8\pi \sin^2 \theta_W}
\frac{m^2_{\mu}}{m^2_{S}}\tan \beta \left(1-\frac{4\alpha}{\pi}
\ln \frac{m_{S}}{m_{\mu}}\right),
\end{equation}
where $m_S$ denotes the superparticle mass in the loop.
It has been pointed out already that this new data on $a_{\mu}$
provides useful 
information on SUSY parameters 
\cite{kane,feng,nath,yama,hisano,ibrahim,ellis,arnowitt,kim,wells1},
e.g. upper bounds on some superparticle masses.
It is also noted that much of the parameter
space of the minimal anomaly-mediated model
can be excluded by the new data when combined with
the constraints from $b\rightarrow s\gamma$ \cite{feng}.
Possible origin of $\Delta a_{\mu}$ other than SUSY is discussed
also in Refs. \cite{nosusy}.

In this paper, we wish to study the implications of
the precisely measured  $a_{\mu}$ 
for  various mediation models of SUSY breaking 
which lead to predictive forms of flavor conserving
soft parameters.
The models studied here include the dilaton/modulus-mediated model
in heterotic string/$M$ theory, no-scale or gaugino-mediated model, 
gauge-mediated model, and also the minimal and deflected
anomaly-mediated models \cite{anomaly,moreanomaly}.
In the subsequent analysis,
we explore the possibility that the deviation 
(\ref{deviation}) is due to the
SUSY contribution to $a_{\mu}$ in these models.
Throughout the analysis, we will assume that soft parameters
(approximately) conserve CP, which may be necessary to
avoid a too large neutron electric dipole moment.
If one takes (\ref{deviation}) as it is,
the corresponding  $2\sigma$ bound on $a^{\rm SUSY}_{\mu}$
would be given by
\beq
\label{range}
10\times 10^{-10} < a^{\rm SUSY}_{\mu}
< 74\times 10^{-10}.
\eeq

The inclusive 
$b \to s \gamma$ process is known to put
strong constraints on the MSSM parameter space.
The leading SUSY contribution to
$b \to s \gamma$ comes from the charged Higgs boson
and chargino mediated diagrams.
The charged Higgs boson diagram contributes constructively,
while the chargino diagram interferes with the SM amplitude
constructively or destructively depending upon the sign
of $\mu$. 
The branching ratio for $b \to s \gamma$  
is obtained by normalizing the hadronic uncertainty with the semileptonic
decay rate \cite{buras}:
\beq
\frac{{Br}(B \to X_s \gamma)}{{Br}(B \to X_c e \bar{\nu} )}
= \frac{|V^*_{ts}V_{tb}|^2}{|V_{cb}|^2}
\frac{6 \alpha_{\rm em}}{\pi f(z)}(|D|^2 + A) F,
\eeq 
where
$f(z) = (1-8 z + 8 z^3 -z^4 -12 z^2 \ln z)$ is the phase space factor
of the semileptonic decay with $z = m^2_{c}/m^2_{b}$,
$F = [1 - 8\alpha_s(m_b)/3\pi]/\kappa(z)$
for the
QCD correction factor
$\kappa(z)\approx 1-2\alpha_s[2.1 (1-z)^2+1.5]/3\pi$
for the semileptonic decay,
and the term $A$ describes the bremsstrahlung corrections
and virtual corrections satisfying the cancellation of 
the IR divergence \cite{chetyrkin}.
The amplitude $D$ is determined  by the Wilson coefficients
at $m_b$ which can be obtained 
by the matching condition at the weak scale and
the subsequent  RG evolution.
We perform the matching 
at the next-to-leading order (NLO)
for the SM contribution, 
while taking the leading order (LO) matching for the MSSM contributions,
i.e.  the charged Higgs and the chargino contributions.
We then perform the RG evolution down to $m_b$ at the NLO to find 
\beq 
D =  C^{(0)}_7 (m_b) + \frac{\alpha_s(m_b)}{4 \pi}
\left( C^{(1)}_7 (m_b) + 
\sum_{i=1,8} r_i C^{(0)}_i (m_b) \right),
\eeq
where we follow the notation of
\cite{buras} and $r_i$ is quoted in Refs. \cite{chetyrkin,kagan}.
Combining the recent CLEO \cite{CLEO} and the ALEPH \cite{ALEPH}
results, one finds the $2\sigma$ constraint \cite{kagan}
\beq
2.18 \times 10^{-4} < Br(B \to X_s \gamma) < 4.10 \times 10 ^{-4},
\eeq 
which will be used to constrain 
the parameter space in our analysis.

For $M_3M_2>0$,
the parameter region of 
$a^{\rm SUSY}_{\mu}>0$ is constrained by the lower
bound on $Br(B\to X_s\gamma)$, while that of $a^{\rm SUSY}_{\mu}
<0$ is constrained by the upper bound.
In this regard, the minimal anomaly mediation model 
is exceptional since it predicts $M_3M_2<0$,
so $a^{\rm SUSY}_{\mu}>0$ is constrained
by the upperbound on $Br(B\to X_s\gamma)$
\cite{feng}. 
It is expected that the constraint from the 
lower bound becomes weaker when
the NLO effects are included \cite{giudice},
while the constraint from the upper bound can become even stronger
\cite{feng}.

About the bounds on Higgs boson and superparticle masses,
we use the LEP limit $m_h>113.5$ GeV \cite{Higgsmass}
and $m_{\tilde{\tau}}>72$ GeV \cite{LEPSUSYWG}.
The other superparticle mass bounds \cite{LEPSUSYWG}
are satisfied in the allowed 
region of the Higgs and stau mass limits except for
the case of deflected anomaly mediation in which the chargino
mass bound $m_{\chi_1^{\pm}}>103$ GeV plays an important role.

In Figs. 1--14,
we identify the parameter space of the model
which can  give  $a_{\mu}^{\rm SUSY}$ in the
$2\sigma$ range (\ref{range}), while
taking into account other experimental constraints,
e.g. $b\rightarrow s\gamma$ and
the collider bounds on superparticle masses.
For dilaton/modulus mediation models in heterotic string/$M$ theory,
$b\to s\gamma$ 
favors $a^{\rm SUSY}_{\mu}$ smaller than
the $-1\sigma$ value ($26\times 10^{-10}$).
It should be remarked that this constraint is from
the lower bound on $Br(B\to X_s\gamma)$, so can be 
relaxed by the NLO SUSY effects \cite{giudice,song}.
The no-scale model is similarly (but less) 
constrained by $b\to s\gamma$.
Gauge mediation models with low messenger scale
can give any $a^{\rm SUSY}_{\mu}$ within the $2\sigma$ range
(\ref{range}). However models with high messenger scale
favor $a^{\rm SUSY}_{\mu}$ below
the central value ($42\times 10^{-10}$).
The minimal anomaly mediation is constained  
by the upper bound on $Br(B\to X_s\gamma)$ 
implying $a^{\rm SUSY}_{\mu}$ smaller than
the central value. When the NLO correction
to the charged Higgs contribution is included,
$b\to s\gamma$ contrains $a^{\rm SUSY}_{\mu}$
more severely \cite{feng}.
Possible value of $a^{\rm SUSY}_{\mu}$ 
in deflected anomaly mediation is severely constrained
by the superparticle mass bounds and
also $b\to s\gamma$, but still there is a small parameter region
which gives right value of $a^{\rm SUSY}_{\mu}$.

To set up the notation, let us consider generic
low energy interactions of the MSSM fields.
They consist of
supersymmetric couplings encoded in the superpotential 
\beq
W=\frac{1}{6} y_{ijk} \Phi_i\Phi_j\Phi_k
-\mu H_1H_2,
\eeq
and also soft supersymmetry breaking terms
which can be written as
\beq
\label{soft}
-{\cal L}_{\rm SB}
=\pm\frac{1}{2}M_a\lambda_a\lambda_a
+\frac{1}{2} m_{ij}^2 \phi_i\phi^*_j
+\frac{1}{6}A_{ijk}y_{ijk}\phi_i\phi_j\phi_k
+B\mu h_1h_2+{\rm h.c.}
\eeq
where $y_{ijk}$ denote the Yukawa coupling constants for the
MSSM superfields $\Phi_i$ which include the quark
superfields, the lepton superfields, and also
the two Higgs doublet superfields $H_1$ and $H_2$.
Here $M_a$ ($a=3,2,1$) stand for  the
$SU(3)\times SU(2)\times U(1)$ gaugino masses,
$m^2_{ij}$ are the soft scalar masses of the scalar components
$\phi_i$ of the MSSM superfields $\Phi_i$,
and $A_{ijk}$ and $B$ are the trilinear and bilinear 
coefficients in the scalar potential.
To follow up the most frequently used convention
for the relative sign of $M_a$ and
$A_{ijk}$, we use different sign conventions of 
$M_a$ for different models:
$+$  for the dilaton/modulus and no-scale (gaugino) mediation models,
$-$ for the gauge and anomaly mediation
models.

Each  mediation mechanism that will be studied
in this paper provides a well-defined prediction for
$M_a$, $A_{ijk}$ and $m^2_{ij}$ at certain high
energy messenger scale.
The predicted high energy parameters can be 
transformed to the low energy values through the standard
renormalization group (RG) analysis.
In this procedure, we assume the minimal particle content
in the observable sector, viz. the MSSM particles.
If there exist more particles with masses
between the messenger scale and the weak scale
and also with sizable gauge or Yukawa couplings
to the MSSM fields, our results would be changed
accordingly.
We also ignore the effects of small Yukawa couplings
of the 1st and 2nd generations in the RG evolution.

The situation for
$\mu$ and $B$ is more involved since
they depend on the details of how the $\mu$-term is generated
as well as on how SUSY breaking is mediated.
In the absence of any definite prediction
for $\mu$ and $B$, normally one trade $\mu$ and $B$ for
$\tan\beta=\langle H_2\rangle/\langle H_1\rangle$ and
$M_Z$ through the condition of radiative electroweak
symmetry breaking, while
leaving ${\rm sign}(\mu)$ undetermined.
Note that $\mu$ and $B$ do not affect the RG running of
other soft parameters, which can be assured by
the dimensional argument and selection rules.

It has been noted that an extensive region of
the soft parameter space gives rise to a scalar potential
with a color or charge breaking minimum or a field
direction along which the potential is unbounded from
below \cite{munoz,casas,abel}. 
For instance, it turns out that
the entire parameter space of the dilaton/modulus
mediation in heterotic string theory and also of
the no-scale mediation give such a potentially dangerous
scalar potential \cite{munoz,casas}.
In this paper, we do {\it not} require that the scalar
potential should have a  phenomenologically
viable {\it global} minimum,
so the model is allowed as long as the scalar potential 
has a {\it local} minimum with correct low energy phenomenology. 

We also do not take into account the cosmological mass
density of the lightest superparticle (LSP) in the MSSM
sector. There are many different scenarios in which
the LSP mass density computed in the framework
of $R$ parity conserving MSSM becomes irrelevant,
e.g. a late time inflation triggered by an MSSM
singlet, $R$-parity violation, or a modulino/gravitino lighter than 
the LSP.

\section{Probing the messengers of supersymmetry breaking}

In this section, we  examine the low energy phenomenology
of various mediation mechanisms yielding flavor conserving
soft parameters. The main purpose is to see which value
of $a^{\rm SUSY}_{\mu}$ can be obtained without 
any conflict to $b\rightarrow s\gamma$ and the collider
bounds on superparticle masses.
The models studied here include the dilaton/modulus-mediated
model in heterotic string/$M$ theory, no-scale or gaugino-mediated
model, gauge-mediated model, and finally the minimal and deflected 
anomaly-mediated models.
For each mediation model,
the parameter regions allowed by  laboratory
tests and also the corresponding value of
$a^{\rm SUSY}_{\mu}$ are summarized  in Figs. 1--14.

\subsection{Dilaton/modulus mediation in perturbative heterotic string
theory}

One possible scheme for flavor-conserving 
soft parameters is 
the dilaton/modulus mediation in the framework of weakly coupled 
heterotic string theory. 
The K\"ahler potential and the gauge kinetic function
of the four-dimensional effective supergravity are given by
\bea
&& K=-\ln (S+S^*)-3\ln (T+T^*)+(T+T^*)^{n_i}\Phi_i\Phi^*_i
\nonumber \\
&& 4\pi f_a=S, 
\eea
where $S$ and $T$ are the dilaton superfield and 
the overall modulus superfield,
respectively, and $n_i$ is the modular weight
of the chiral matter superfields $\Phi_i$.
If all the MSSM superfields have the modular
weight $n_i=-1$, one finds (at the unification
scale $M_{\rm GUT}$) \cite{string}
\beq
\label{dilaton}
M_a=\sqrt{3}M_{\rm aux},
\quad
m^2_{ij}=|M_{\rm aux}|^2\delta_{ij},
\quad
A_{ijk}=-\sqrt{3}M_{\rm aux},
\eeq
where $M_{\rm aux}=m_{3/2}\sin\theta$ for the Goldstino
angle $\theta$ which is defined as
$\tan\theta=F_S/F_T$.
Here we assume that $F_S/F_T$ is real to avoid 
a too large neutron electric dipole moment.
The above relations  can receive
string-loop or supergravity-loop corrections \cite{stringloop,sugraloop} 
as well as higher order sigma-model 
corrections \cite{choi}. In weakly coupled heterotic string limit,
loop corrections are suppressed by
$g^2_{\rm GUT}/8\pi^2$, so can be safely ignored
for our purpose.
Also at least in orbifold compactification models,
there is no sigma-model correction at string tree level.

In fact, the gauge coupling unification scale
$M_{\rm GUT}$ predicted within
the weakly coupled heterotic string theory
is bigger than the phenomenologically favored value 
$2\times 10^{16}$ GeV by about one order of magnitude.
One attractive way to avoid this difficulty is to go
to the strong coupling limit \cite{witten}, i.e. the Horava-Witten heterotic
$M$-theory \cite{witten1}, 
which will be analyzed in the subsequent
discussion.
Here we simply assume that $M_{\rm GUT}$ can be lowered 
down to $2\times 10^{16}$ GeV by some stringy effects,
while keeping the boundary conditions of (\ref{dilaton}) valid.
Another potential problem of the boundary condition (\ref{dilaton})
is that the resulting scalar potential has a color or charge breaking
minimum or has a field direction which is  unbounded from below
\cite{casas}.
We do not take this as a serious problem as long as
there exists a local minimum of the potential yielding
correct low energy phenomenology.  

Some phenomenological consequences of (\ref{dilaton})  has been
studied in \cite{string1}.
Here we perform a detailed numerical analysis of the low energy
phenomenology of the boundary condition (\ref{dilaton}) at $M_{\rm GUT}$,
including the SUSY contributions
to $a_{\mu}$ and $b\rightarrow s\gamma$.
As usual, we trade $\mu$ and $B$
for $\tan\beta$ and $M_Z$.
With this prescription, the dilaton/modulus mediation
in perturbative heterotic string theory is described by three
input parameters,
\beq
M_{\rm aux}, \quad \tan\beta, \quad  {\rm sign}(\mu).
\eeq 
The results of our analysis are depicted in Fig. 1
including the contour plot on the plane of $(M_{\rm aux},
\tan\beta)$ with $\mu>0$.
Fig. 1 shows that
$b\to s\gamma$ favors $a^{\rm SUSY}_{\mu}$ {\it smaller} than
the $-1\sigma$ value ($26\times 10^{-10}$).
This constraint from $b\to s\gamma$  
is expected to be relaxed when the 
NLO SUSY corrections to $Br(B\to X_s\gamma)$ are properly
taken into account \cite{song}.

\bigskip

\subsection{Dilaton/modulus mediation in heterotic $M$ theory}

It has been pointed out by Witten that the
correct value of $M_{\rm GUT}$ can be naturally obtained
in compactified heterotic $M$ theory which corresponds to
the strong coupling limit of heterotic $E_8\times E_8$
string theory \cite{witten}.
At energy scales below the eleven-dimensional Planck scale,
the theory is described by an eleven-dimensional supergravity
on a manifold with boundary where the two $E_8$ gauge multiplets
are confined on the two ten-dimensional boundaries \cite{witten1}.
The compactified 
heterotic $M$ theory involves two geometric moduli,
the eleventh length ($\pi\rho$) and the volume
($V$) of six dimensional
internal space.
In four-dimensional effective supergravity \cite{mtheory,mtheory1},
these two moduli define the scalar components of the chiral superfields
$S$ and $T$,
\beq
{\rm Re}(S)=\frac{V}{(4\pi)^{2/3}\kappa^{4/3}},
\quad
{\rm Re}(T)=\frac{V^{1/3}\pi\rho}{(4\pi\gamma)^{1/3}
\kappa^{2/3}},
\eeq
where $\kappa^2$ is the eleven-dimensional gravitational
coupling constant and $\gamma=
\frac{1}{6}\sum_{IJK} C_{IJK}$  for the intersection
numbers $C_{IJK}=\int \omega_I\wedge\omega_J\wedge\omega_K$
of the integer (1,1) cohomology basis \{$\omega_I$\}.
Here the superfields $S$ and $T$ are normalized  through
the periodicity of their axion components: ${\rm Im}(S)\equiv
{\rm Im}(S)+1$ and ${\rm Im}(T)\equiv
{\rm Im}(T)+1$.

Four-dimensional couplings and scales
can be expressed in terms of ${\rm Re}(S)$, ${\rm Re}(T)$
and $\kappa$,
yielding the relations \cite{banks,choi1}
\bea
\label{mscale}
&& \frac{M^2_{\rm P}}{M^2_{\rm GUT}}
=4\pi \gamma^{1/3} {\rm Re}(S) {\rm Re}(T),
\nonumber \\
&& \frac{4\pi}{g^2_{\rm GUT}}={\rm Re}(S)+\alpha {\rm Re}(T),
\eea
where $M_{\rm P}\approx 2.5 \times 10^{18}$ GeV and
$g^2_{\rm GUT}\approx 0.5$ are the four-dimensional
Planck scale and gauge coupling constant, respectively,
and
$\alpha$ is a model-dependent (positive) rational number
which is generically of order unity.
Putting $M_{\rm GUT}\approx 2\times 10^{16}$ GeV, one then finds
the following vacuum expectation values (VEVs)
of moduli in heterotic $M$-theory: 
\beq
\label{modulivalue}
\langle {\rm Re}(S)\rangle={\cal O}\left(\frac{4\pi}{g^2_{\rm GUT}}\right),
\quad \quad
\langle {\rm Re}(T)\rangle={\cal O}\left(\frac{4\pi}{g^2_{\rm GUT}}\right).
\eeq

It has been noted that the four-dimensional effective supergravity
of heterotic $M$-theory can be expanded in powers of
$1/\pi (S+S^*)$ and $1/\pi (T+T^*)$ \cite{mtheory}. 
At leading order in this expansion,
the K\"ahler potential and gauge kinetic function are given by
\cite{mtheory1}
\bea
&& K=-\ln (S+S^*)-3\ln (T+T^*)
+\left(\frac{3}{T+T^*}+
\frac{\alpha}{S+S^*}\right)\Phi_i\Phi_i^*,
\nonumber \\
&& 4\pi f_a= S+\alpha T.
\eea
In fact, holomorphy and the axion periodicity implies that
any correction to $f_a$ is suppressed by $e^{-2\pi S}$ or
$e^{-2\pi T}$, so absolutely negligible for the 
moduli VEVs of Eq. (\ref{modulivalue}).
The K\"ahler potential can receive corrections
which are higher order in $1/\pi (S+S^*)$ or
$1/\pi (T+T^*)$. For the moduli VEVs (\ref{modulivalue}),
the effects of such higher order corrections are suppressed by
$g^2_{\rm GUT}/8\pi^2$, so can be ignored also for our purpose.
With this observation, one finds the following form of
soft parameters in heterotic $M$-theory (again
at $M_{\rm GUT}$) when SUSY breaking is
mediated by the $F$-components of $S$ and $T$,
\bea
\label{msoft}
M_a&=&\sqrt{3}m_{3/2}\left(
\frac{1}{1+\epsilon}\sin\theta +\frac{\epsilon}{\sqrt{3}(1+
\epsilon)}\cos\theta\right),
\nonumber \\
A_{ijk}&=&-\sqrt{3}m_{3/2}\left(\frac{3-2\epsilon}{3+\epsilon}
\sin\theta+\frac{\sqrt{3}\epsilon}{3+\epsilon}\cos\theta\right),
\nonumber \\
m^2_{ij}&=&|m_{3/2}|^2\delta_{ij}\left(
1-\frac{3}{(3+\epsilon)^2}\left\{
\epsilon(6+\epsilon)\sin^2\theta
\right.\right.
\nonumber \\
&&
\left.\left.
+(3+2\epsilon)\cos^2\theta
-2\sqrt{3}\epsilon \cos\theta\sin\theta\right\}\right)
\eea
where $\theta$ is the Goldstino angle and
$$
\epsilon=\alpha (T+T^*)/(S+S^*).
$$

The above results express $M_a$, $A_{ijk}$ and $m^2_{ij}$
in terms of three unknown parameters $m_{3/2}$, $\sin\theta$ and
$\epsilon$.
Once the $\mu$ and $B$ are traded  for $\tan\beta$ and $M_Z$
through the condition of radiative electroweak symmetry breaking,
the dilaton/modulus mediation in heterotic $M$-theory is described
by five input parameters,
\beq
m_{3/2}, \quad \sin\theta, \quad \epsilon,\quad \tan\beta,
\quad {\rm sign}(\mu),
\eeq
so not more predictive
than the minimal supergravity model for instance. 
However in heterotic $M$-theory, the value of $\epsilon$
is severely constrained, which allows the results of
(\ref{msoft}) become  more predictive.
For instance, the hidden gauge coupling is given by
$4\pi/g^2_{\rm H}=(1-\epsilon){\rm Re}(S)$, so
it is required that $0<\epsilon<1$.
Inspecting (\ref{mscale}), one also finds
that $\epsilon$ can not be significantly smaller than the unity.

Here we consider two different values $\epsilon=
0.5, \, 0.8$, and    
examine the allowed value of
$a^{\rm SUSY}_{\mu}$.
The results of our analysis are depicted in Figs. 2--5
for $(\epsilon,\tan\beta)=(0.5,10), (0.8, 10),
(0.5, 30), (0.8, 30)$.
These figures show that 
$b\to s\gamma$ favors
$a_{\mu}^{\rm SUSY}$ smaller than the $-1\sigma$ value 
($26\times 10^{-10}$).
Again this constraint is expected to be
relaxed when the NLO SUSY corrections to $Br(B\to X_s\gamma)$ are included.
We note that $a^{\rm SUSY}_{\mu}$
for $\tan\beta\lesssim 10$ is significantly constrained
by other laboratory bounds also, e.g. the lightest Higgs mass bound.

\subsection{No-scale or gaugino mediation}

It has been known for a long time that no-scale supergravity
model with non-minimal gauge kinetic function provides an interesting
form of flavor conserving soft terms \cite{noscale}.
For instance, one can consider the no-scale K\"ahler potential
together with the simplest non-minimal gauge kinetic functions:
\beq
K=-3\ln (T+T^*-\Phi_i\Phi^*_i),
\quad
4\pi f_a=T,
\eeq
which  give rise to 
\beq
\label{noscale}
M_a=M_{\rm aux},
\quad
m^2_{ij}=0, \quad 
A_{ijk}=0.
\eeq
at the messenger scale  which is close to the unification scale.
Recently it has been noticed that such no-scale
boundary condition can naturally emerge in the framework
of brane models in which SUSY is broken on
a hidden brane in higher dimensional spacetime \cite{gaugino}.
The MSSM matter fields are assumed to be confined on a visible
brane. However gauge multiplets propagate  in bulk
and so couple directly to SUSY breaking on hidden brane.
Extra-dimensional locality then assures that the soft
parameters of the MSSM matter fields vanish, i.e. $m^2_{ij}=A_{ijk}
=0$, at the compactification scale $M_c$ of the extra dimension, while
nonzero gaugino masses are allowed, leading to the name of
``gaugino mediation'' \cite{gaugino}.

In gaugino-mediated model, 
the compactification scale $M_c$ is a model-dependent free parameter.
If gaugino masses are universal at $M_c=M_{\rm GUT}$, 
it is rather difficult that the LSP is a neutral particle 
\cite{gaugino1,noscale1}.
One can avoid this difficulty either by assuming
$M_c>M_{\rm GUT}$ or non-universal gaugino masses
\cite{gaugino1,noscale1}.
However a neutral LSP is not mandatory.
For instance, a charged LSP is allowed if
$R$-parity is broken or the model includes a modulino lighter than the
charged LSP.
Here we assume that the no-scale boundary condition (\ref{noscale}) 
is given at $M_{\rm GUT}=2\times 10^{16}$ GeV and examine 
the resulting SUSY contribution to the muon's anomalous magnetic moment.
About $\mu$ and $B$, in gaugino-mediated model,
it is rather natural that $B=0$ at $M_{\rm GUT}$.
However in generic no-scale supergravity model,
$B$ can be a free parameter, and then
the no-scale mediation 
is described by three input parameters,
\beq
M_{\rm aux}, \quad  \tan\beta, \quad  {\rm sign}(\mu).
\nonumber 
\eeq

The results of our numerical analysis  are summarized in Fig. 6
which is somewhat similar to Fig. 1, i.e. the case of dilaton/modulus
mediation in heterotic string theory.
The analysis of $b\rightarrow s\gamma$ for the 
no-scale boundary condition (\ref{noscale}) has been performed recently
in Ref. \cite{noscale2}.
It should be remarked also that the scalar potential 
resulting from (\ref{noscale})
has a color or charge breaking minimum
or a field direction along which the potential is unbounded from below
\cite{munoz}.
As we mentioned, we do not take this as a serious difficulty
as long as the potential has a phenomenologically viable local minimum.

\subsection{Gauge mediation}

The gauge-mediated SUSY breaking (GMSB) models also
provide a quite predictive form of flavor conserving
soft parameters \cite{gauge}.
In GMSB models,  SUSY breaking is transmitted via
the SM gauge interactions of $N$ flavors of
messenger superfields $\Psi_i,\Psi^c_i$  which form 
a  vector-like representation
of the SM gauge group, e.g. $N({\bf 5}+{\bf \bar{5}})$ of
$SU(5)$. 
Then the resulting soft terms are determined by
the gauge quantum numbers, so automatically conserve
the flavors.
The messenger fields are coupled to a gauge singlet Goldstino
superfield $X$
through the superpotential
\begin{equation}
 W = \lambda_i X \Psi_i\Psi^c_i. 
\end{equation}

When $X$ aquires a VEV for both
its scalar and $F$ components,
the superpotential $W$ induces the messenger spectrum which is not
supersymmetric.
Integrating out the messenger fields then give rise to
the following MSSM soft parameters
at the messenger scale $M\approx \lambda_i\langle X\rangle$:
\bea
&& M_a = N\frac{\alpha_a(M)}{4\pi} \Lambda,
\nonumber \\
&& m^2_{ij}=2N\delta_{ij}\sum_{a} C^i_a \left(\alpha_a(M)\over {4\pi}\right)^2
\Lambda^2,
\nonumber \\
&& A_{ijk}=0,
\eea
where 
$\alpha_a$ ($a=3,2,1$) are
the GUT-normalized gauge coupling constants of $SU(3)_c\times
SU(2)_L\times U(1)_Y$,  $C^i_a$ is the GUT-normalized
quadratic Casimir invariant of the matter field $\Phi_i$, and
$\Lambda\approx \langle F_X\rangle/\langle X\rangle$.

In fact, the trilinear couplings $A_{ijk}$ at the messenger
scale $M$ receive nonzero contribution at two-loop,
however we can safely ignore them since
they are further suppressed by the loop factor
compared to other soft masses with mass dimension one.
Again $\mu$ and $B$ can be traded for
$\tan\beta$ through the radiative electroweak symmetry breaking.
A distinctive feature of GMSB  is that a wide range
of the messenger scale $M$ is allowed, e.g. from $\Lambda$
to much higher scale around $10^{15}$ GeV.
Then the GMSB model is described by
five input parameters,
\beq
M, \quad \Lambda,  \quad \tan\beta, \quad N, \quad 
{\rm sign}(\mu).
\nonumber
\eeq

Low energy phenomenology of GMSB models,
including 
$b\rightarrow s\gamma$ and the anomalous muon magnetic moment
$a_{\mu}$, has been studied before \cite{wells,gauge1}.
Here we examine the allowed value of
$a^{\rm SUSY}_{\mu}$ for the cases of
$(N,M)=(1, 10^6), (1, 10^{10}), (1, 10^{15}),
(5, 10^6), (5, 10^{10}), (5, 10^{15})$,
where the messenger scale $M$ is given in the GeV unit.
The results for $\mu>0$
are depicted in Figs. 7--12  
which show that models with lower
$M$ have a better prospect for 
$a^{\rm SUSY}_{\mu}$ bigger than the central value
($42\times 10^{-10}$).
In particular, gauge-mediated
models with $M\sim 10^6$ GeV can give
any $a^{\rm SUSY}_{\mu}$ within the $2\sigma$ bound
(\ref{range}).
For very high $M\sim 10^{15}$ GeV, $b\to s\gamma$
constrains $a^{\rm SUSY}_{\mu}$ as in
no-scale or dilaton/modulus mediation model.

\subsection{Minimal anomaly mediation}

Anomaly mediation assumes that SUSY breaking in the hidden
sector  is transmitted to the MSSM fields {\it only} through
the auxiliary component $u$ of the off-shell supergravity multiplet.
In the Weyl-compensator formulation, 
$u$ corresponds to the $F$-component of the Weyl compensator
superfield $\phi$ in appropriate gauge.
The couplings of $\phi$ to generic matter multiplets are
determined  by the super-Weyl invariance.
Therefore at classical level, $\phi$ is coupled to the MSSM fields
only through  dimensionful (supersymmetric) couplings, e.g
the bare $\mu$ parameter or the coefficients of non-renormalizable
terms in the superpotential.
However quantum radiative effects induce non-trivial scale
dependence of dimensionless couplings, so
non-trivial couplings of $\phi$ also.
Since $M_a$, $A_{ijk}$, $m^2_{ij}$ are all 
associated with dimensionless supersymmetric couplings, 
viz. the gauge couplings
$g_a$ for $M_a$, the wave function renormalization factor
$Z_i$ for $A_{ijk}$ and $m^2_{ij}$, these soft parameters are determined
entirely by the running behavior of $g_a$ and $Z_i$
in pure anomaly-mediated scenario. 
More explicitly, from the super-Weyl invariant effective lagrangian,
\beq
\int d^4
\theta \left({Z}_i
(\mu/\sqrt{\phi\phi^*}) \Phi_i^*\Phi_i
+\frac{1}{8}g^{-2}_a(\mu/\sqrt{\phi\phi^*}) V^aD\bar{D}^2DV^a+...\right),
\eeq
one finds the following {\it pure} anomaly-mediated
soft parameters
\bea
&& \tilde{M}_a=\frac{1}{2}
g_a^2\left(\frac{dg_a^{-2}}{d \ln\mu }\right)\frac{F_{\phi}}{\phi}
=-\frac{b_a\alpha_a}{4\pi} M_{\rm aux},
\nonumber \\
&& \tilde{A}_{ijk}=-\frac{1}{2}\left(\frac{d \ln Z_i}{d\ln\mu}
+\frac{d \ln Z_j}{d\ln\mu}+\frac{d \ln Z_k}{d\ln\mu}\right)
\frac{F_{\phi}}{\phi}
=\frac{1}{2}(\gamma_i+\gamma_j+\gamma_k)M_{\rm aux},
\nonumber \\
&& \tilde{m}^2_{ij}=-\frac{1}{4}\delta_{ij}
\left(\frac{d^2 \ln Z_i}{d(\ln\mu)^2}
\right)\left|
\frac{F_{\phi}}{\phi}\right|^2=
-\frac{\dot{\gamma}_i}{4}|M_{\rm aux}|^2\delta_{ij},
\eea
where $V^a$ are the real superfields for gauge multiplets,
$b_a=(3,-1,-33/5)$ ($a=3,2,1$) are the one-loop beta function
coefficients for $SU(3)_c\times SU(2)_L\times U(1)_Y$
in the GUT normalization.

The expressions of pure anomaly-mediated soft parameters 
are RG-invariant, so are valid at 
arbitrary energy scale.
Therefore the low energy soft parameters are completely fixed
by the low energy values of these couplings and an overall scale
$M_{\rm aux}$.
However, as can be seen easily,  pure anomaly-mediated
scenario is simply excluded because it predicts
that sleptons have negative mass-squared.
 So any phenomenologically viable model of anomaly mediation
should involve a mechanism to solve the tachyonic
slepton problem. One possibility is to introduce a universal
positive mass-squared to all soft scalar masses at some
high energy scale, e.g. at $M_{\rm GUT}$, which
defines the minimal anomaly-mediated model:
\bea
&&
M_a=\tilde{M}_a, \quad
A_{ijk}=\tilde{A}_{ijk},
\nonumber \\
&& m^2_{ij}(M_{\rm GUT})= \tilde{m}^2_{ij}(M_{\rm GUT}) +m_0^2
\delta_{ij}.
\eea

After trading $\mu$ and $B$ for $\tan\beta$ and $M_Z$,
the minimal anomaly mediation can be parameterized by
four input parameters,
\beq
M_{\rm aux}, \quad m_0, \quad 
\tan\beta, \quad {\rm sign}(\mu).
\nonumber
\eeq  
Phenomenological aspects of the minimal anomaly-mediated model
have been studied in detail in Ref.\cite{anomaly1}.
It has been noted also that the very recent measurement of
the anomalous magnetic moment of the muon disfavors
the minimal anomaly-mediated model when combined with the
constraint from $b\rightarrow s\gamma$ \cite{feng}.
Unlike other models, the minimal anomaly mediation model
predicts $M_3M_2<0$, so the parameter region of
$a^{\rm SUSY}_{\mu}>0$ is constrained by
the upper bound on $Br(B\to X_s\gamma)$.

Our results depicted in Fig. 13  are similar to 
\cite{feng}. However 
in our case, $b\rightarrow s\gamma$ provides less severe
constraint because we use the LO matching condition for
the SUSY contributions to $b\to s\gamma$.
Including the NLO
charged Higgs contribution \cite{song}
makes the constraint from $b\rightarrow s\gamma$ stronger
as in Ref. \cite{feng}.

\subsection{Deflected anomaly mediation}

There is an interesting modification of
the pure anomaly mediation which cures the tachyonic slepton
with $M_3M_2>0$ \cite{danomaly1,danomaly}.
The parameter region of $a^{\rm SUSY}_{\mu}>0$
in such model may not be severely constrained by
$b\to s\gamma$.
The so-called deflected anomaly mediation
is a kind of hybrid between
anomaly mediation and gauge mediation, but still all
SUSY breaking effects originate from $F_{\phi}$.
The model contains a light
singlet $X$ which describes a flat direction in supersymmetric limit
as well as $N$ flavors of gauge-charged messengers 
$\Psi_i,\Psi^c_i$ which are coupled to $X$ in the superpotential 
\beq
W=\lambda_i X\Psi_i\Psi^c_i.
\eeq
If the VEV of $X$ is determined by the SUSY breaking effects of $F_{\phi}$,
{\it not} by SUSY conserving dynamics,
one has
\beq
\frac{F_X}{X}=\rho \frac{F_{\phi}}{\phi}
\eeq
where $\rho$ depends on the details of how
$X$ is stabilized, but $\rho\neq 1$ in general.
For instance, in case that $X$ is  stabilized  by
the Coleman-Weinberg mechanism, one finds
$\rho={\cal O}(1/16\pi^2)$ \cite{moreanomaly,danomaly}.

At energy scales below $M\approx \lambda_i\langle
X\rangle$,
the heavy thresholds effects of
$\Psi_i,\Psi^c_i$ make all soft parameters
to leave the RG trajectory of pure anomaly mediation.
We then have
\begin{eqnarray}
&& M_a(M) = -(b_a - N(1-\rho)){\alpha(M)\over 4\pi} M_{\rm aux}, 
\nonumber  \\
&& A_{ijk}(M) =\tilde{A}_{ijk} (M), 
\nonumber \\
&& m_{ij}^2(M) =\tilde{m}^2_{ij}(M)
    -2N(1-\rho)\delta_{ij}\sum_a{C_a^i}\left(
{\alpha_a(M)\over 4\pi}\right)^2|M_{\rm aux}|^2\,,
\end{eqnarray}
where $b_a=(3,-1,-33/5)$ ($a=3,2,1$), 
and $\tilde{A}_{ijk}$, $\tilde{m}^2_{ij}$ are
the pure anomaly-mediated soft parameters
in the MSSM framework.
Then the deflected anomaly mediation
is described by  six input parameters,
\beq
M_{\rm aux}, \quad M, \quad \rho, \quad \tan\beta,
\quad N, \quad {\rm sign}(\mu).
\eeq

For numerical analysis, we take $\rho\approx 0$ which corresponds
to the case that $X$ is stabilized by the Coleman-Weinberg
mechanism \cite{danomaly}.
The results for $(N,\tan\beta)=
(6, 30)$
are depicted in Fig. 14.
Possible value of $a^{\rm SUSY}_{\mu}$ in this model  is severely
constrained by the chargino, stau, and lightest Higgs mass bounds
as well as  by $b\to s\gamma$, however
there is still a small parameter region which 
provides right value of $a_{\rm SUSY}$.

\section{conclusion}

The recent BNL  measurement of the muon's anomalous magnetic
moment $a_{\mu}\equiv (g_{\mu}-2)/2$ may indeed be a sign of new physics.
In this paper, we examined the possibility that
the deviation $\Delta a_{\mu}$ from
the SM value is due to the  SUSY contribution 
in the framework of
various mediation models of SUSY breaking
which give rise to predictive
flavor conserving soft parameters at high energy scale.
The studied models include
the dilaton/modulus mediation in heterotic string/$M$ theory,
no-scale or gaugino mediation, gauge mediation, and also the
minimal and deflected anomaly mediation models.
For each model, we obtain the range of $a^{\rm SUSY}_{\mu}$
allowed by other laboratory constraints, e.g.
$b\rightarrow s\gamma$ and the bounds on superparticle
masses, together with the corresponding parameter
region of the model.

For dilaton/modulus mediation models in heterotic string/$M$ theory,
the lower bound on $Br(B\to X_s\gamma)$ favors
$a^{\rm SUSY}_{\mu}$ smaller than the $-1\sigma$ bound
($26\times 10^{-10}$).
No-scale model is similarly (but less) 
constrained by $b\to s\gamma$.
Gauge mediation models with low messenger scale
can give any $a^{\rm SUSY}_{\mu}$ within the $2\sigma$ bound
(\ref{range}).
However when the messenger scale is very high,
the allowed value of $a^{\rm SUSY}_{\mu}$ is
constrained by $b\to s\gamma$ as in no-scale or dilaton/modulus
mediation models.
The possible value of $a^{\rm SUSY}_{\mu}>0$ 
in minimal anomaly mediation is constrained
to be smaller than the central value
($42\times 10^{-10}$) by the upper bound on $Br(B\to X_s\gamma)$.
Deflected anomaly mediation is severely constrained
by the superparticle mass bounds and also
by $b\to s\gamma$, however there is still a small parameter
region which gives  right value of $a^{\rm SUSY}_{\mu}$. 
Our analysis uses the LO matching condition for the SUSY
contributions to $b\to s\gamma$. 
It is expected that more involved analysis
including the NLO SUSY effects \cite{giudice} makes
the constraint from the lower bound on $Br(B\to X_s \gamma)$
weaker, while making the constraint from the upper bound stronger.

While this paper is in completion,  there appeared 
several papers which have some overlap with our work.
Ref. \cite{feng} contains a discussion of the minimal
anomaly-mediated model, which agrees with our result,
Refs. \cite{nath},\cite{yama} contain
some discussion of gauge-mediated model, 
and the dilaton-mediated and gauge-mediated models
are discussed in Ref. \cite{wells1} also.

\bigskip

{\bf Acknowledgments}:
We thank Jonathan L. Feng, Konstantin T. Matchev,
James D. Wells and Pyungwon Ko for useful discussions.
This work is supported by the BK21 project of the
Ministry of Education, KRF Grant No. 2000-015-DP0080,
KOSEF Grant No. 2000-1-11100-001-1, 
and KOSEF through the CHEP of KNU.

\begin{figure}
\begin{center}
\epsfig{file=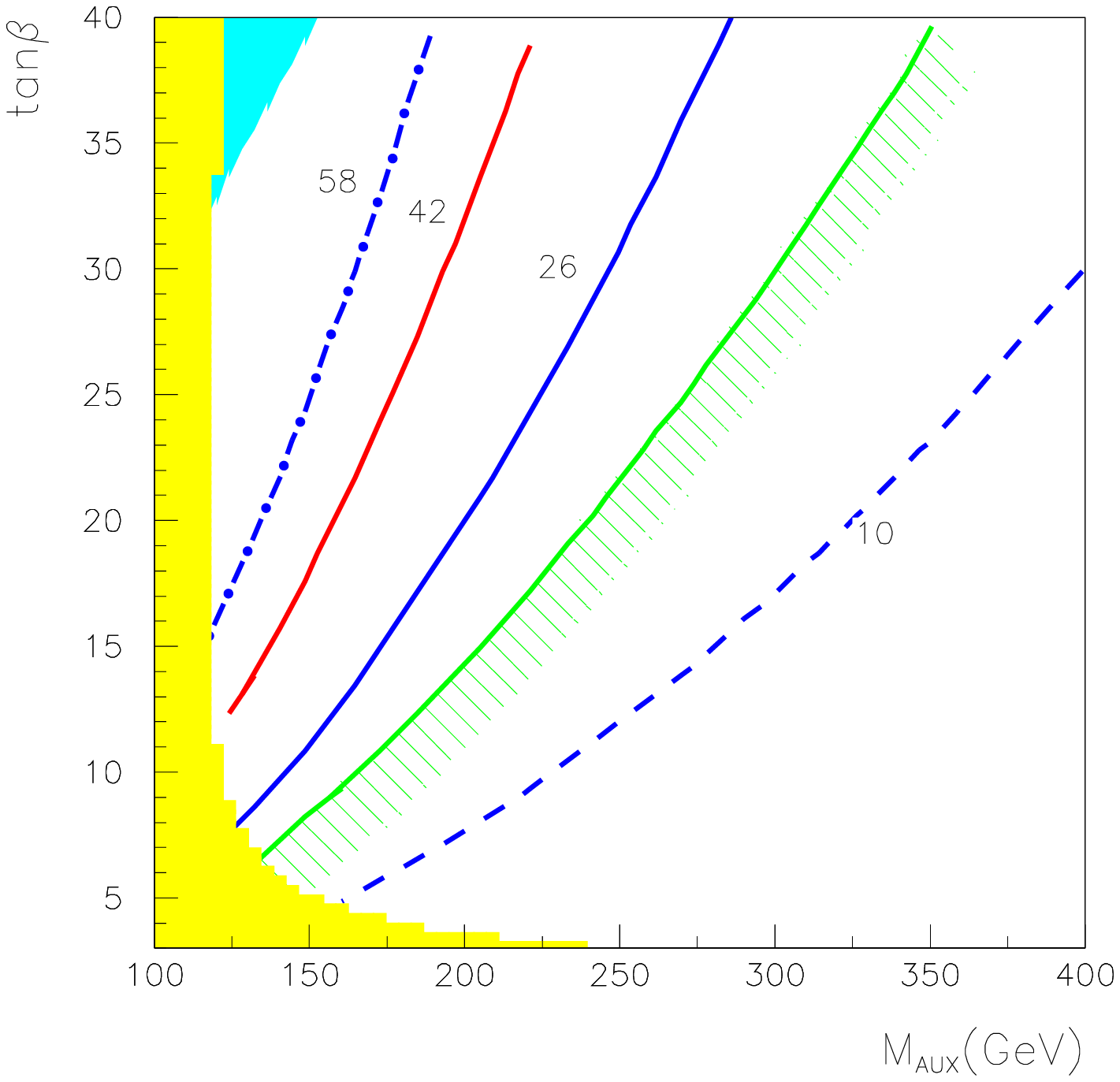, bbllx=60, bblly=60, bburx=480, bbury=466}
\end{center}
\caption{Contour plot  on the plane of  ($M_{\rm aux},\tan\beta$)
in the dilaton/modulus mediation model of heterotic 
string theory.
Yellow and cyan regions represent the parameter spaces 
forbidden by the lightest higgs mass bound and the stau mass bound, 
respectively.
The red line represents  the central value of $a_\mu^{\rm SUSY}$
($42\times 10^{-10}$) and
the blue dash-dotted, solid, dashed lines stand  
for the $+1\sigma$, $-1\sigma$,$-2\sigma$ values of 
$a_\mu^{\rm SUSY}$, respectively.
The green solid line corresponds to the contour of
the $2\sigma$ lower bound $Br(B\to X_s\gamma)=2.18\times 10^{-4}$ 
obtained from the SUSY LO calculation
and the area below the line is the allowed region.
}
\end{figure}

\begin{figure}
\begin{center}
\epsfig{file=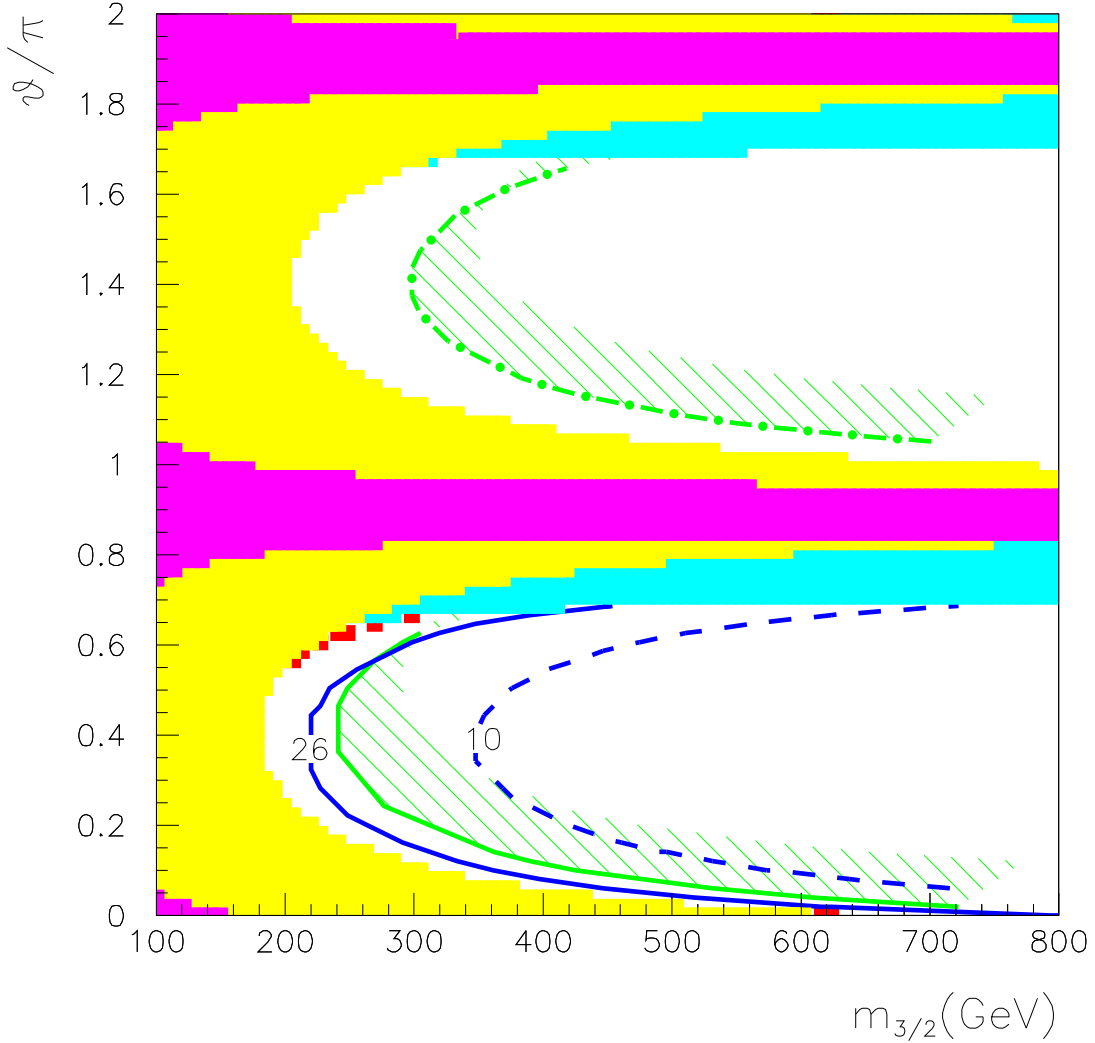, bbllx=60, bblly=60, bburx=480, bbury=466}
\end{center}
\caption{Contour plot on the plane of ($m_{3/2},\theta$) 
in the dilaton/modulus mediation model of heterotic 
$M$ theory with $\epsilon=0.5$ and $\tan\beta=10$.
Yello, purple and cyan regions represent the parameter space 
forbidden by the lightest higgs mass bound, the electroweak
symmetry breaking condition, and the stau
mass bound, respectively.
The blue solid(dashed) line is for the $-1\sigma$($-2\sigma$) value of 
$a_\mu^{\rm SUSY}$ and the red colored region represents the area in which 
$a_\mu^{\rm SUSY}$ is bigger than the central value ($42\times 10^{-10}$).
For $\theta >\pi$,  $a_\mu^{\rm SUSY}$ is negative 
due to the sign flip of the gaugino masses. 
The green solid and green dash-dotted lines stand for
the $2\sigma$ lower and upper bounds on $Br(B\to X_s\gamma)$
and the hashed sides of the lines are the allowed region.
}
\end{figure}

\begin{figure}
\begin{center}
\epsfig{file=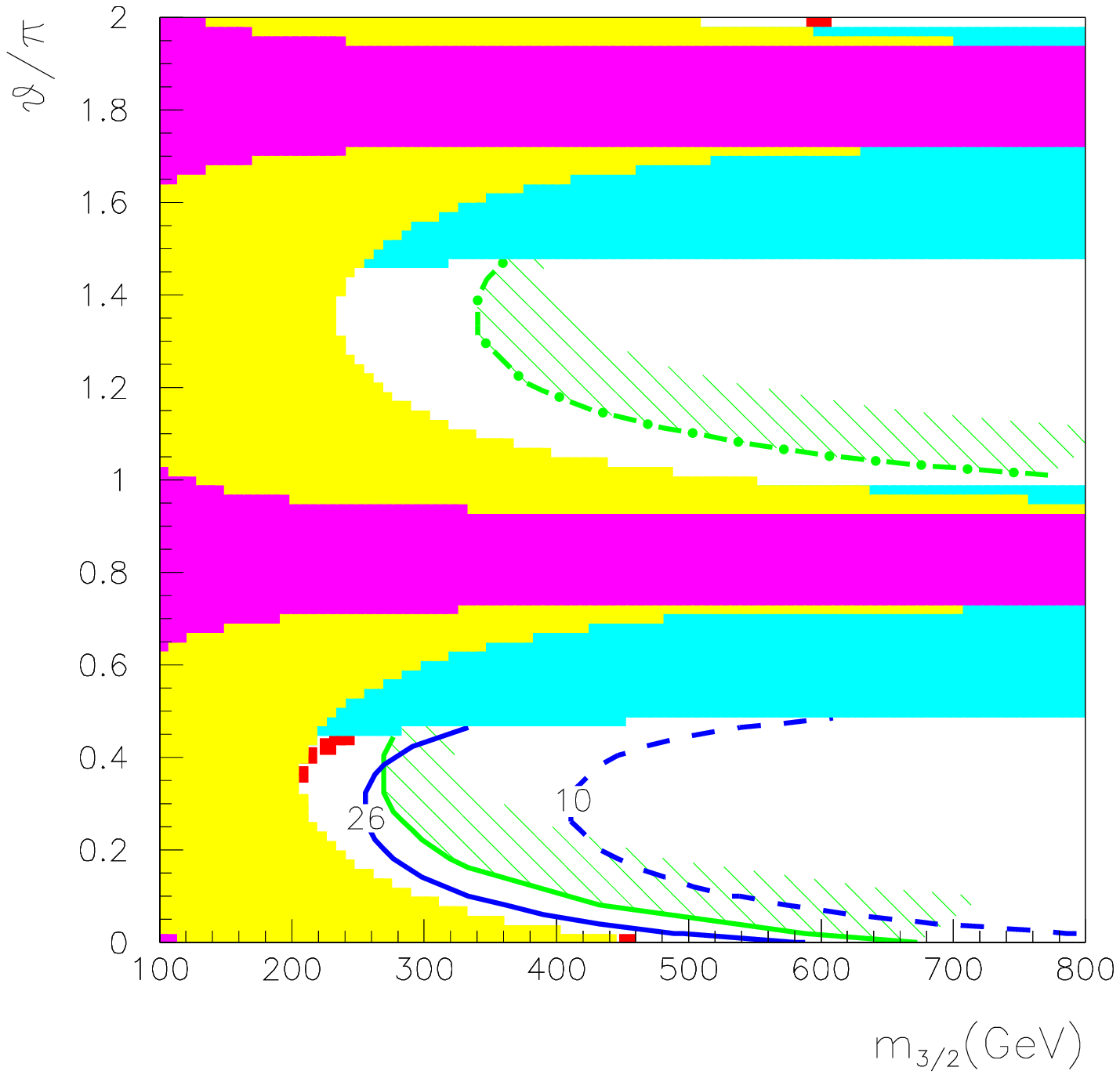, bbllx=60, bblly=60, bburx=480, bbury=466}
\end{center}
\caption{Contour plot on the plane of ($m_{3/2},\theta$)  in the
dilaton/modulus mediation model of  heterotic 
$M$ theory with $\epsilon=0.8$ and $\tan\beta=10$.
}
\end{figure}

\begin{figure}
\begin{center}
\epsfig{file=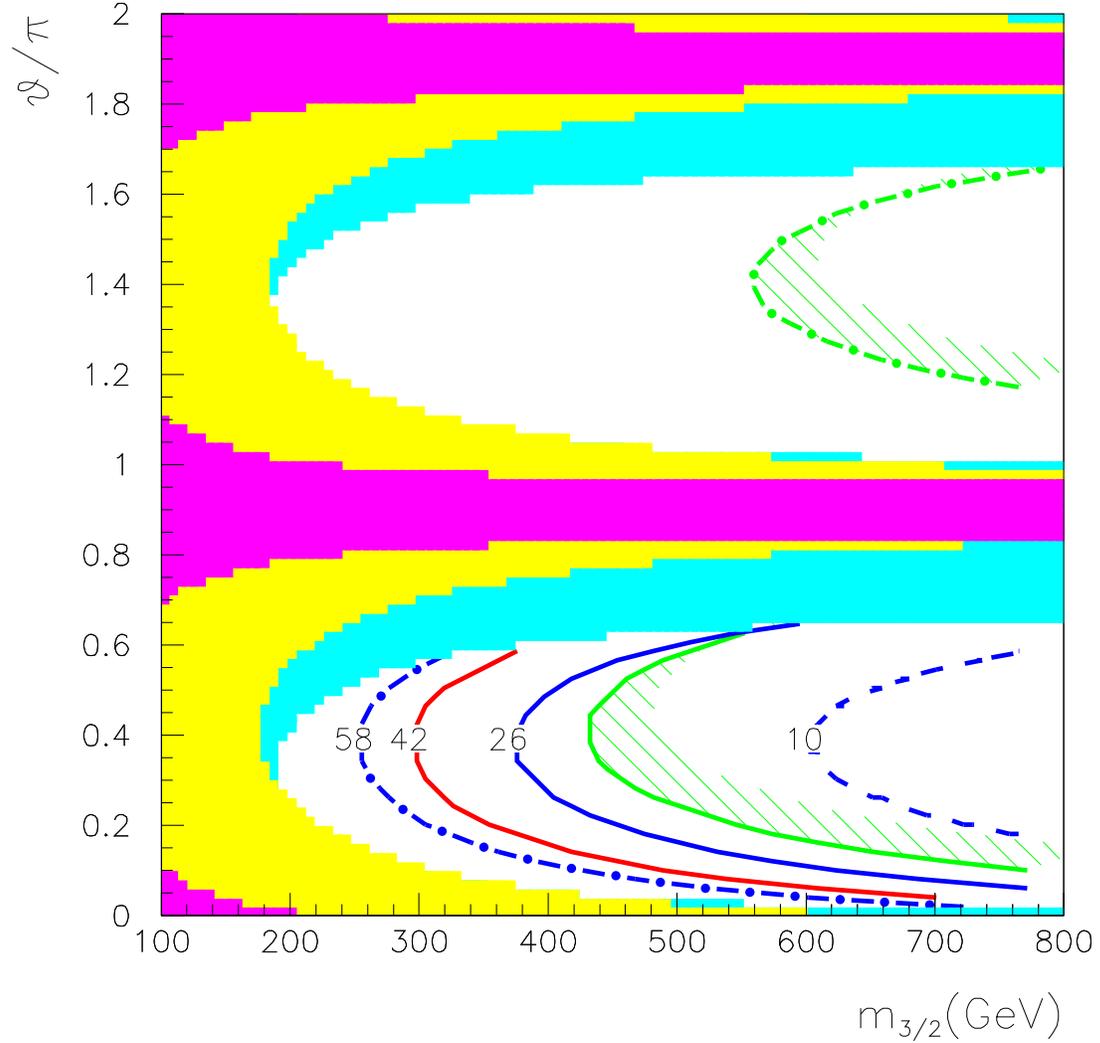, bbllx=60, bblly=60, bburx=480, bbury=466}
\end{center}
\caption{Contour plot on the plane of ($m_{3/2},\theta$) 
in the dilaton/modulus mediation model of  heterotic 
$M$ theory with $\epsilon=0.5$ and $\tan\beta=30$.
The red line is for the centeral value of $a_\mu^{\rm SUSY}$,
and the blue dash-dotted, solid, dashed lines stand 
for the $+1\sigma$, $-1\sigma$, $-2\sigma$ values of 
$a_\mu^{\rm SUSY}$, respectively.
}
\end{figure}

\begin{figure}
\begin{center}
\epsfig{file=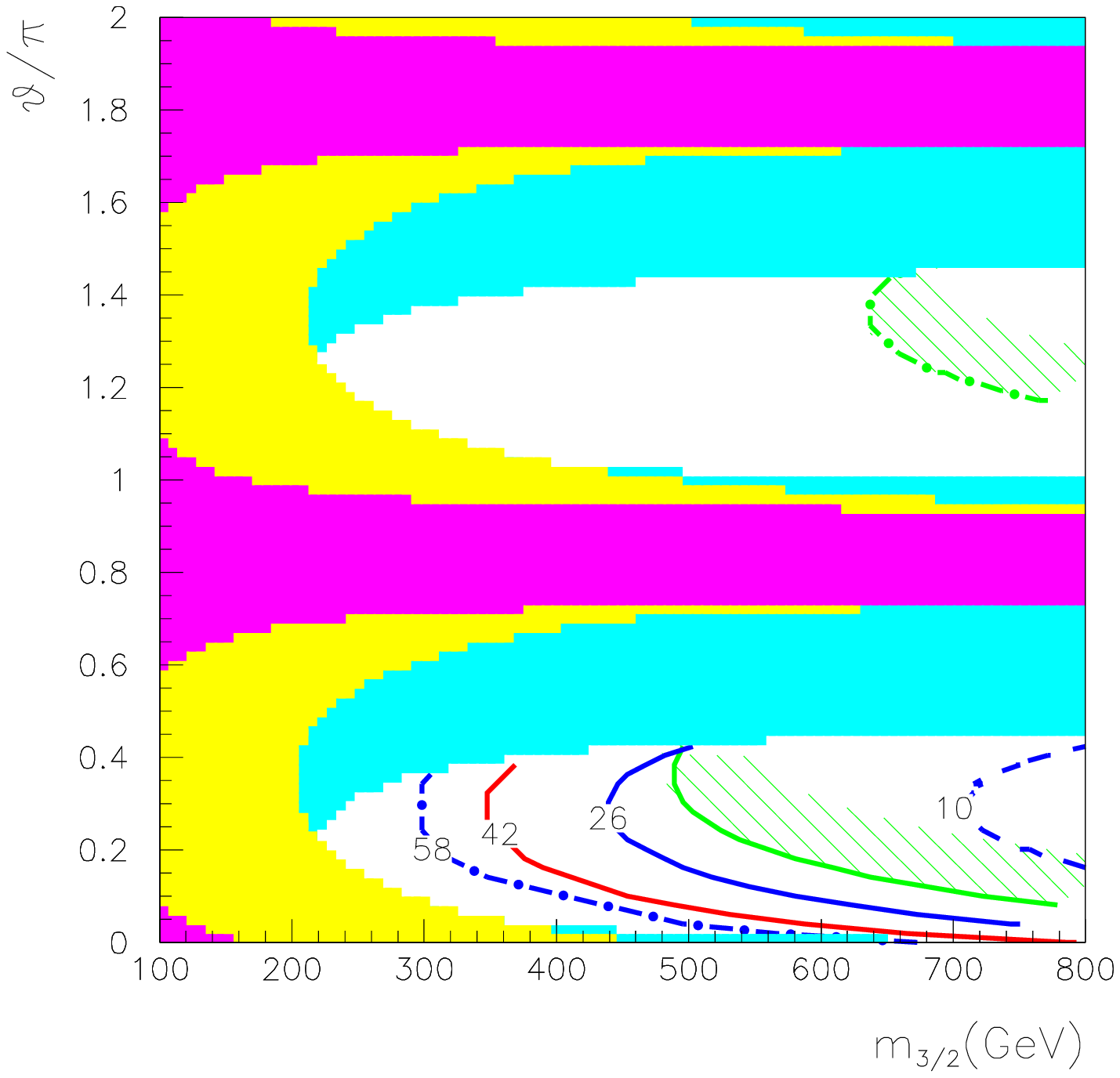, bbllx=60, bblly=60, bburx=480, bbury=466}
\end{center}
\caption{Contour plot on the plane of ($m_{3/2},\theta$)  in the
dilaton/modulus mediation model of  heterotic 
$M$ theory with $\epsilon=0.8$ and $\tan\beta=30$.
}
\end{figure}

\begin{figure}
\begin{center}
\epsfig{file=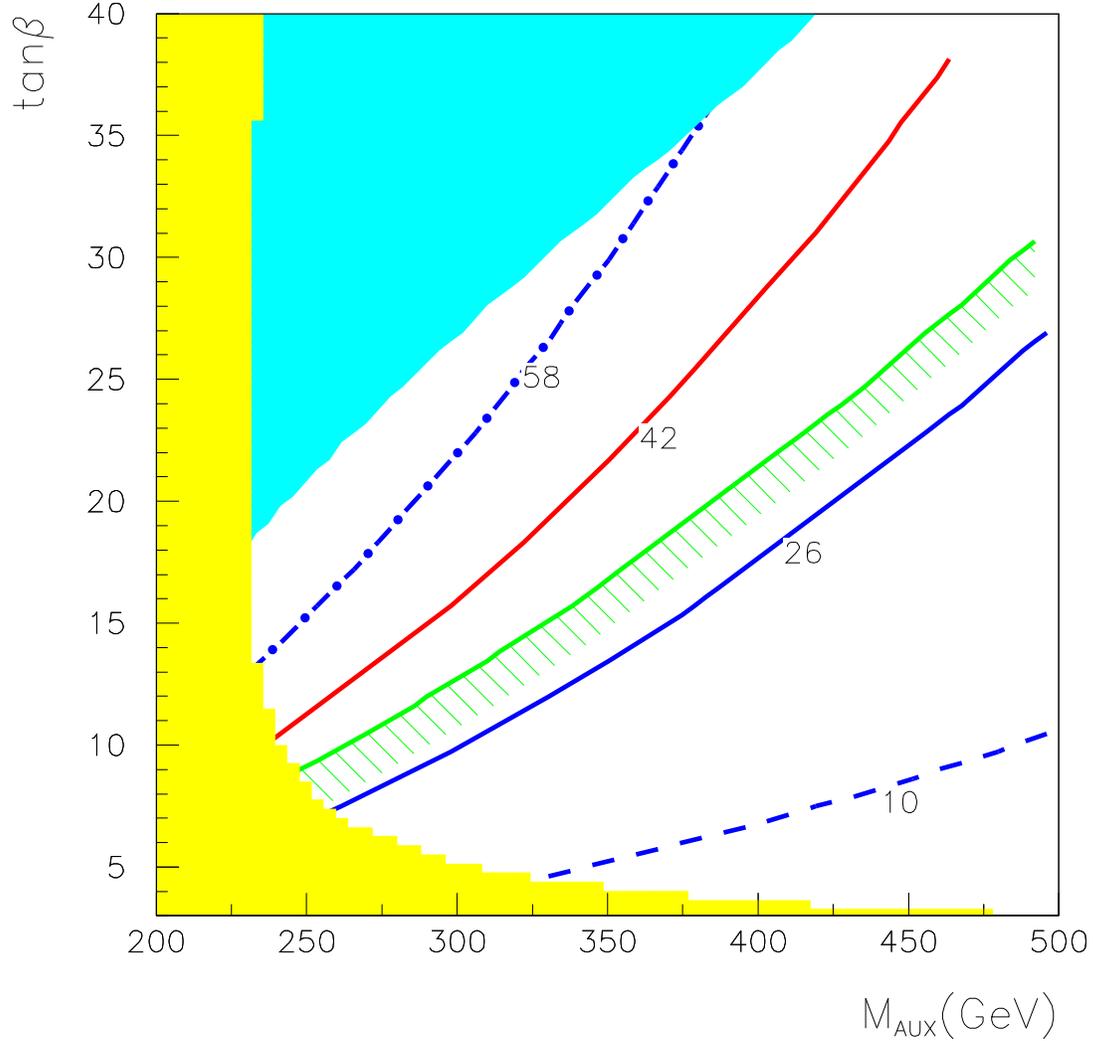, bbllx=60, bblly=60, bburx=480, bbury=466}
\end{center}
\caption{Contour plot on the plane of ($M_{\rm aux},\tan\beta$)
in no-scale model.
Yellow and cyan regions represent the parameter space 
forbidden by the lightest higgs mass bound and the stau mass bound, 
respectively.
The red line is for the centeral value of $a_\mu^{\rm SUSY}$
($42\times 10^{-10}$) and
the blue dash-dotted, solid, dashed lines stand 
for the $+1\sigma$, $-1\sigma$, $-2\sigma$ values of 
$a_\mu^{\rm SUSY}$, respectively.
The green solid line corresponds to  the contour of
the $2\sigma$ lower bound $Br(B\to X_s\gamma)=2.18\times 10^{-4}$
and the area below the line is the allowed region.
}
\end{figure}

\begin{figure}
\begin{center}
\epsfig{file=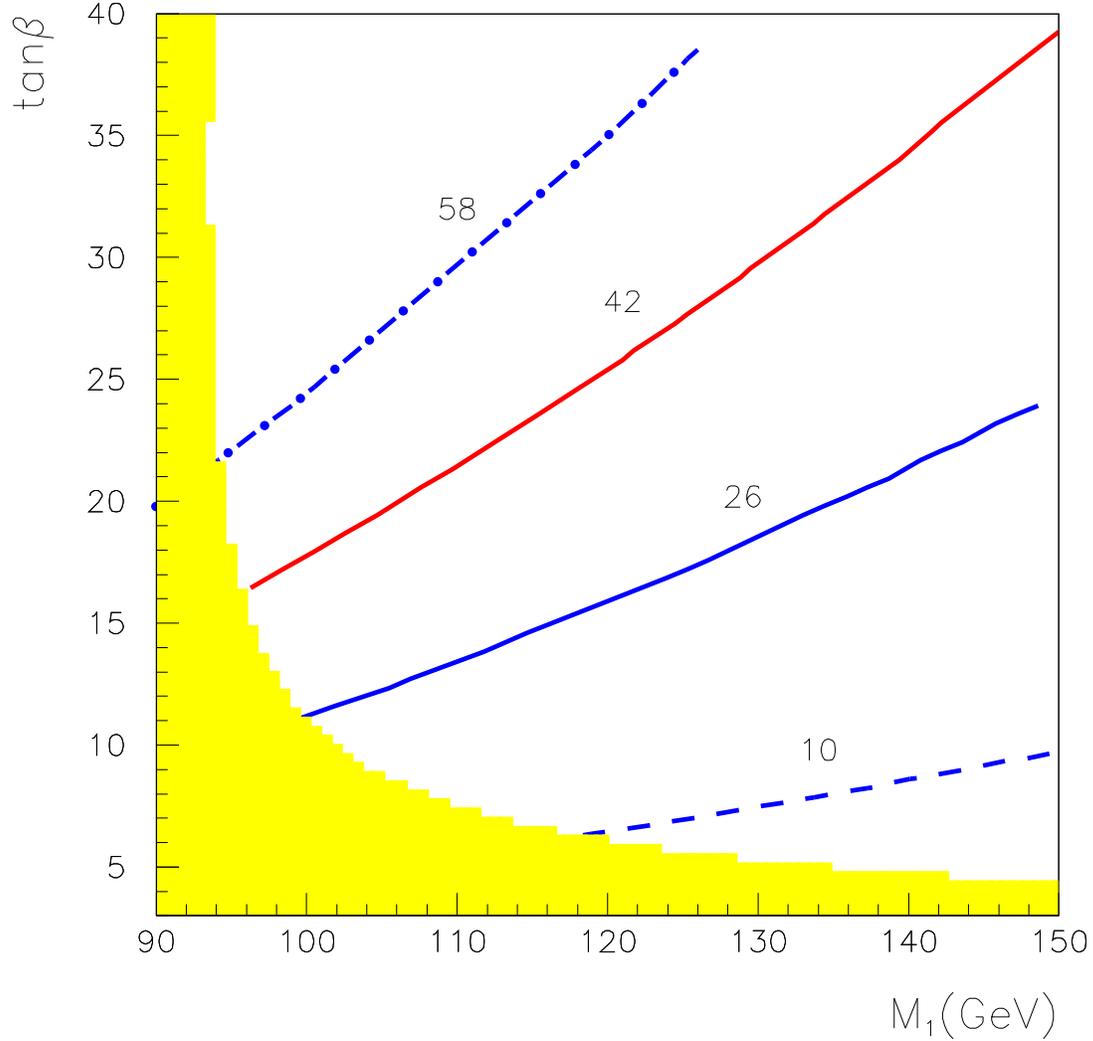, bbllx=60, bblly=60, bburx=480, bbury=466}
\end{center}
\caption{Contour plot on  the plane of ($M_1,\tan\beta$)  in the 
gauge-mediated supersymmetry breaking (GMSB) model
with $N=1$ and $M=10^6$ GeV.
Yellow region represents the parameter space 
forbidden by the lightest higgs mass bound.
The red line is for the centeral value of $a_\mu^{\rm SUSY}$
($42\times 10^{-10}$) and
the blue dash-dotted, solid, dashed lines stand
for the $+1\sigma$, $-1\sigma$, $-2\sigma$ values of 
$a_\mu^{\rm SUSY}$, respectively.
The whole parameter space shown in this figure is allowed
by the constraints of $Br(B\to X_s\gamma)$.
}
\end{figure}

\begin{figure}
\begin{center}
\epsfig{file=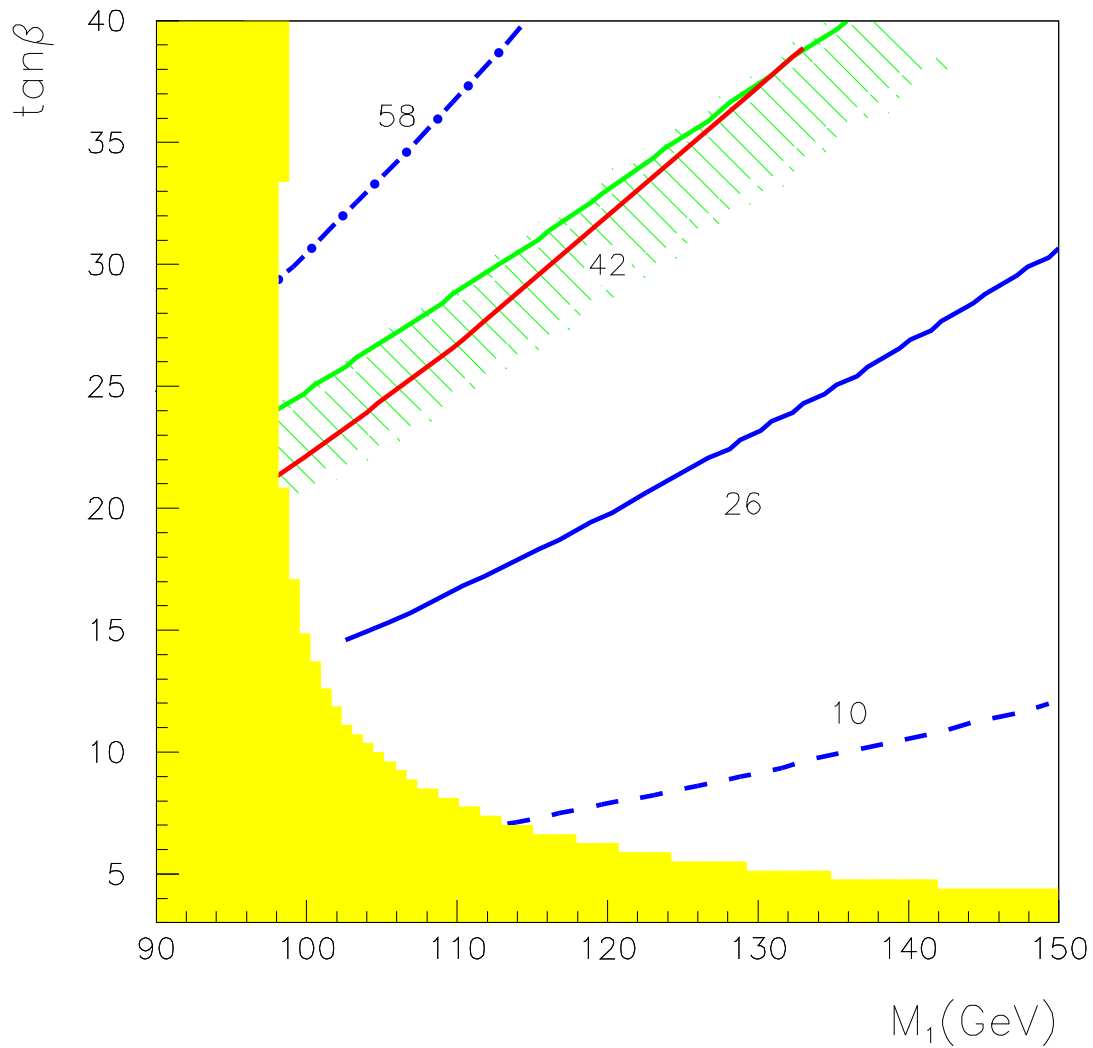, bbllx=60, bblly=60, bburx=480, bbury=466}
\end{center}
\caption{Contour plot on the plane of ($M_1,\tan\beta$) in the GMSB model
with $N=1$ and $M=10^{10}$ GeV.
The green solid line corresponds to  the contour of
the $2\sigma$ lower bound $Br(B\to X_s\gamma)=2.18\times 10^{-4}$
and the area below the line is the allowed region.
}
\end{figure}

\begin{figure}
\begin{center}
\epsfig{file=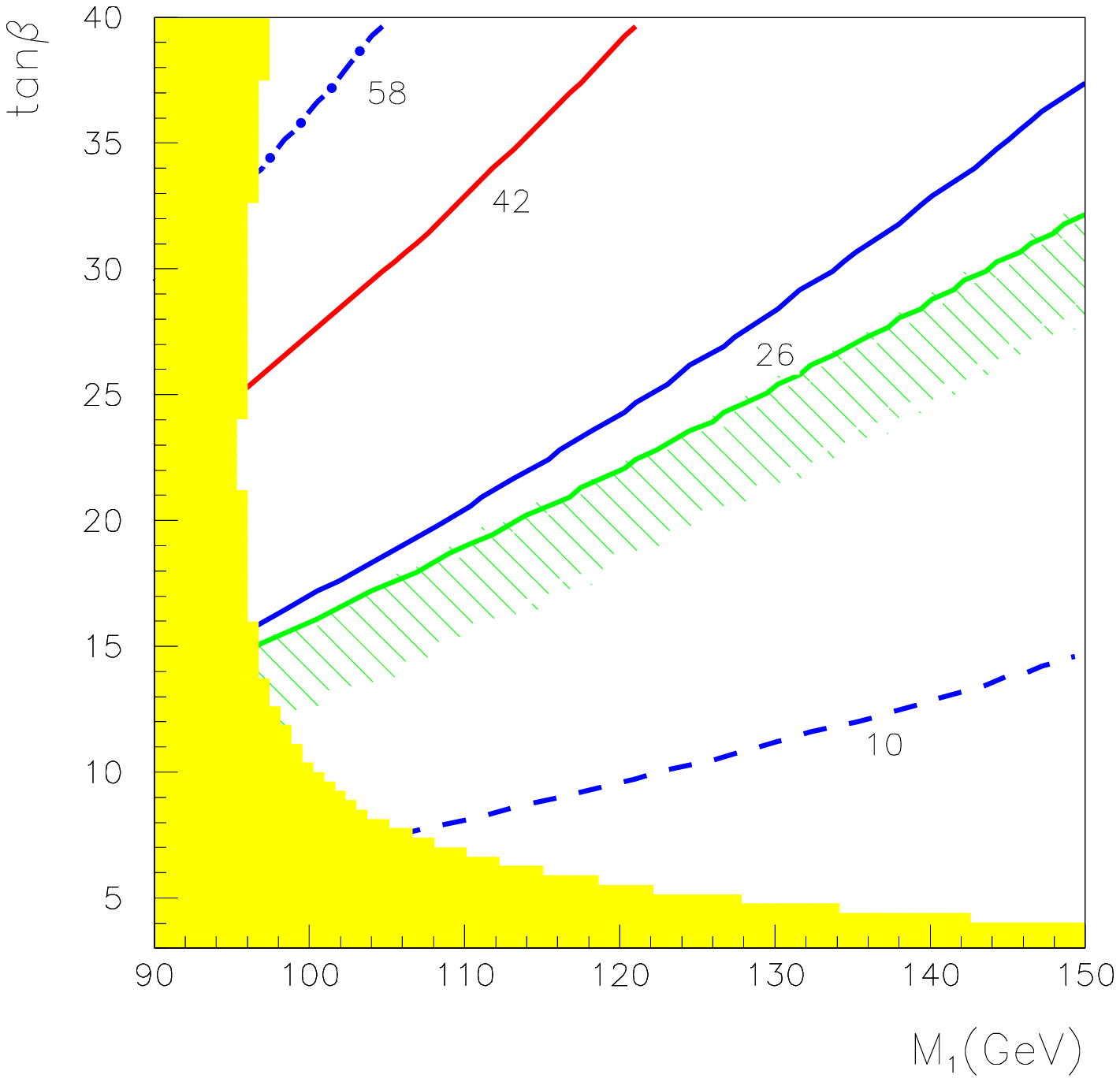, bbllx=60, bblly=60, bburx=480, bbury=466}
\end{center}
\caption{Contour plot on the plane of ($M_1,\tan\beta$)  in the GMSB model
with $N=1$ and $M=10^{15}$ GeV.
}
\end{figure}

\begin{figure}
\begin{center}
\epsfig{file=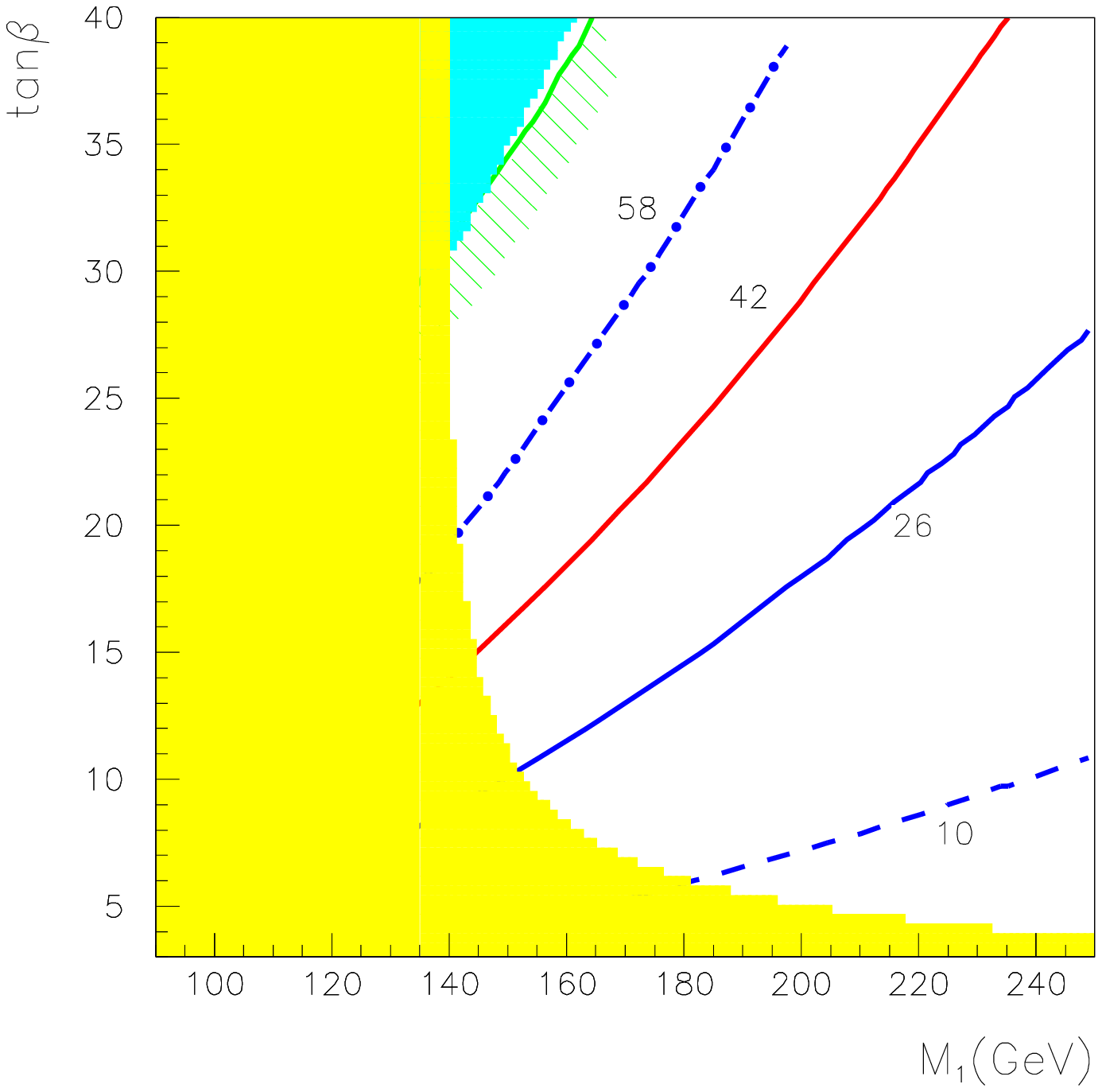, bbllx=60, bblly=60, bburx=480, bbury=466}
\end{center}
\caption{Contour plot on the plane of ($M_1,\tan\beta$) in the GMSB model
with $N=5$ and $M=10^{6}$ GeV.
}
\end{figure}

\begin{figure}
\begin{center}
\epsfig{file=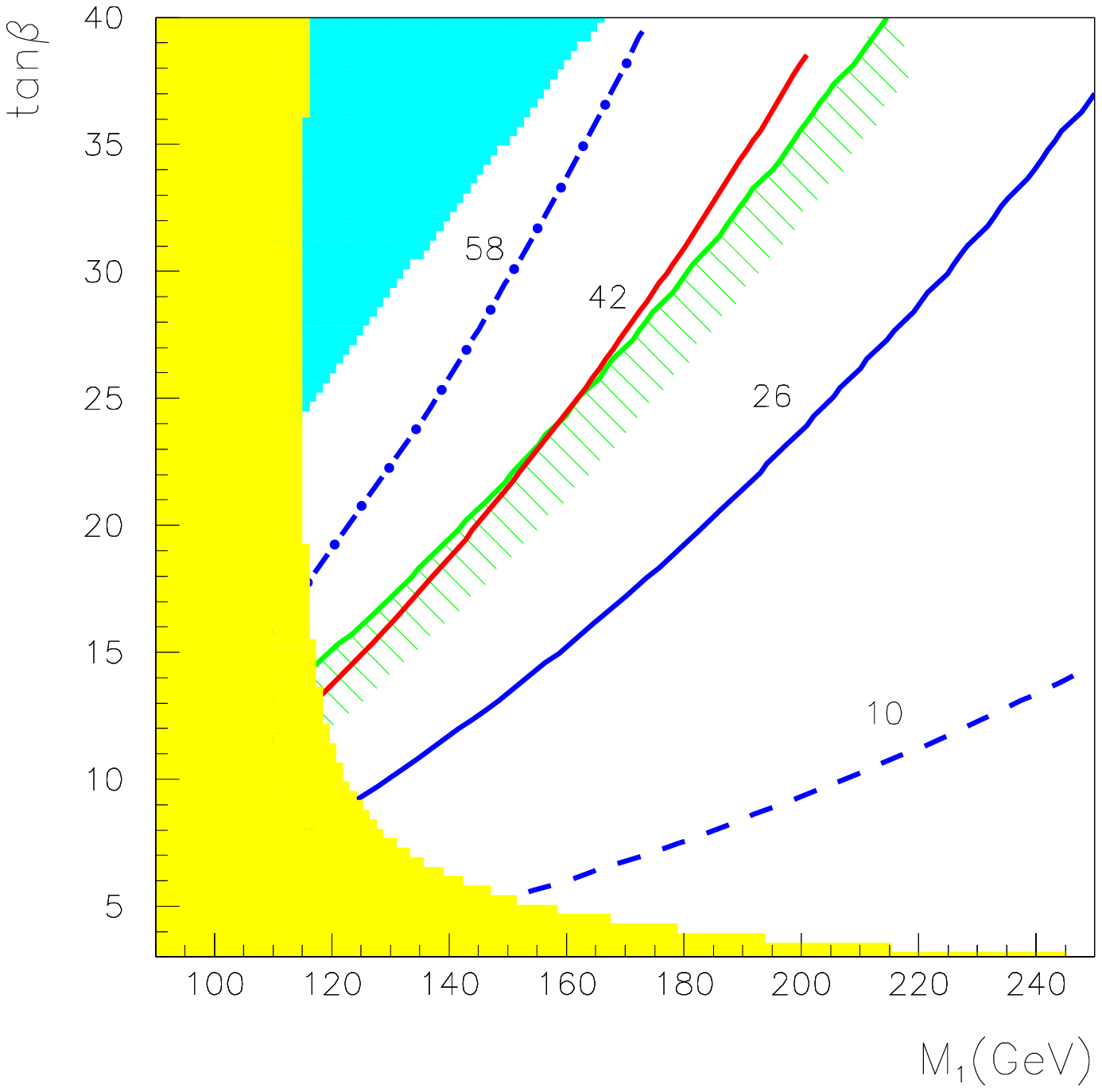, bbllx=60, bblly=60, bburx=480, bbury=466}
\end{center}
\caption{Contour plot on the plane of ($M_1,\tan\beta$)  in the GMSB model
with $N=5$ and $M=10^{10}$ GeV.
}
\end{figure}

\begin{figure}
\begin{center}
\epsfig{file=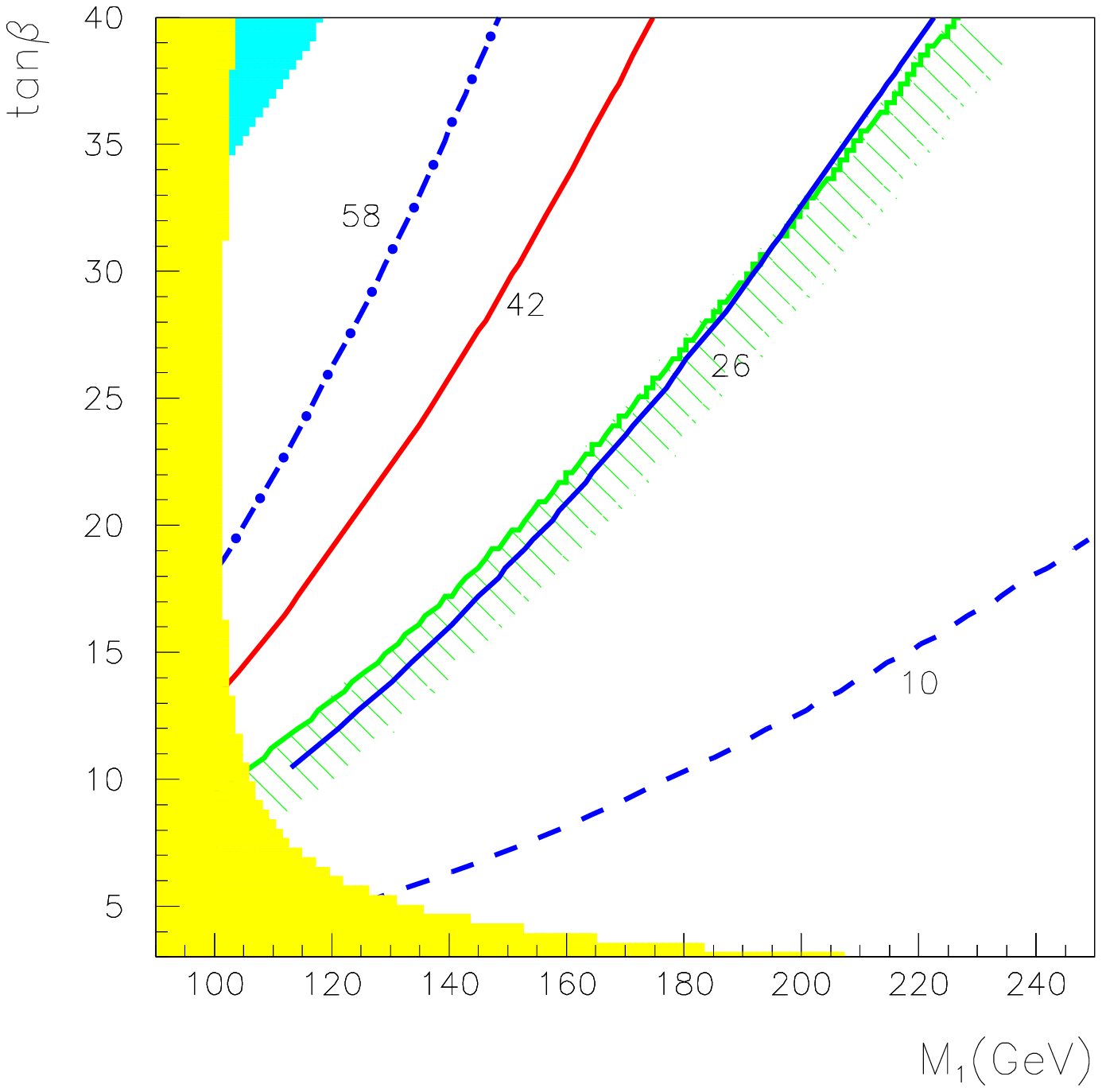, bbllx=60, bblly=60, bburx=480, bbury=466}
\end{center}
\caption{Contour plot on the plane of ($M_1,\tan\beta$)  in the GMSB model
with $N=5$ and $M=10^{15}$ GeV.
}
\end{figure}

\begin{figure}
\begin{center}
\epsfig{file=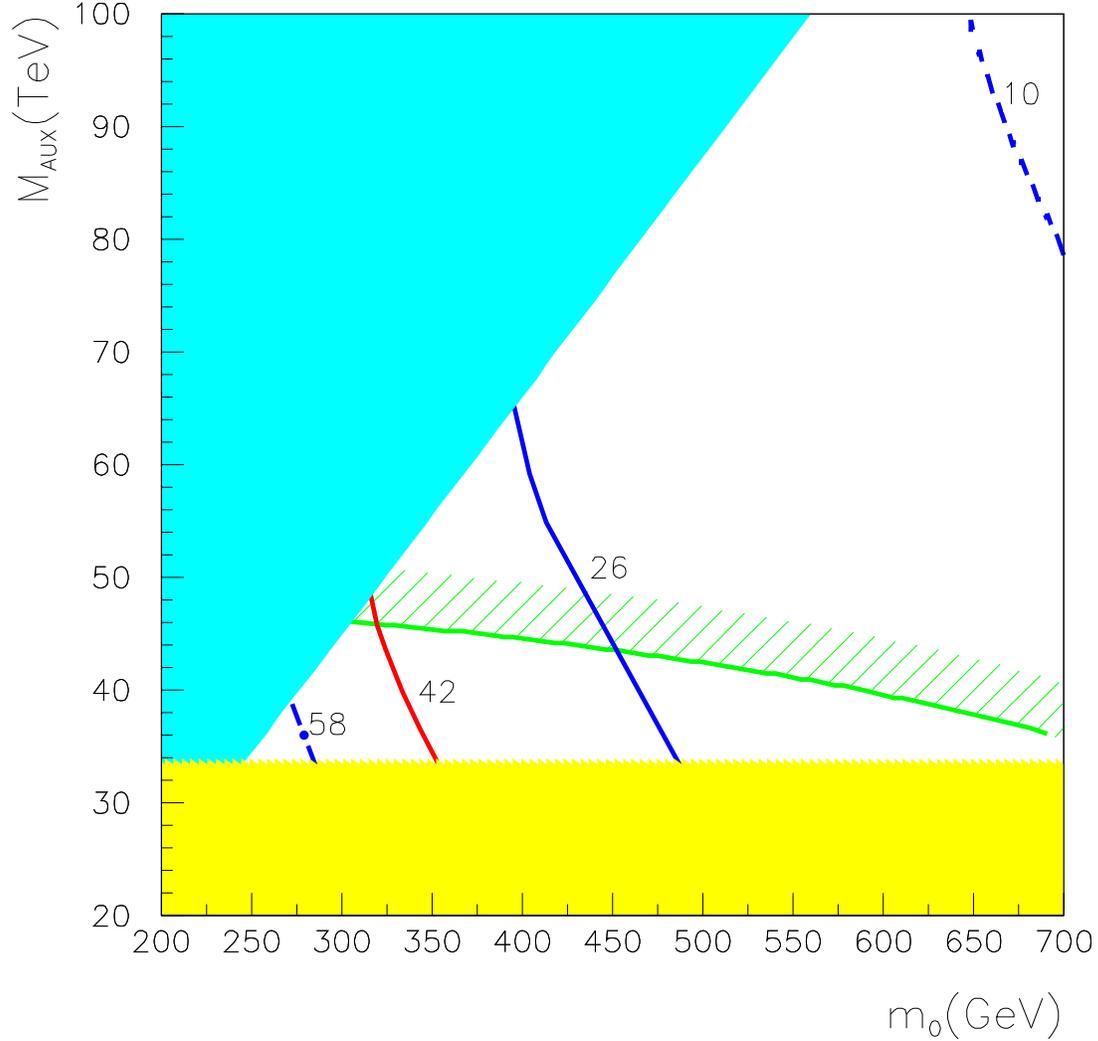, bbllx=60, bblly=60, bburx=480, bbury=466}
\end{center}
\caption{Contour plot on the plane of ($m_0,M_{\rm aux}$)  in the minimal
anomaly-mediated model with $\tan\beta=30$.
Yellow and cyan regions represent the parameter space 
forbidden by the chargino and  stau mass bounds, respectively.
The red line is for the centeral value of $a_\mu^{\rm SUSY}$
($42\times 10^{-10}$) and
the blue dash-dotted, solid, dashed lines stand
for the $+1\sigma$, $-1\sigma$, $-2\sigma$ values of 
$a_\mu^{\rm SUSY}$, respectively.
The green solid line corresponds to the contour of
the $2\sigma$ lower bound $Br(B\to X_s\gamma)=2.18\times 10^{-4}$
and the area above the line is the allowed region.
}
\end{figure}

\begin{figure}
\begin{center}
\epsfig{file=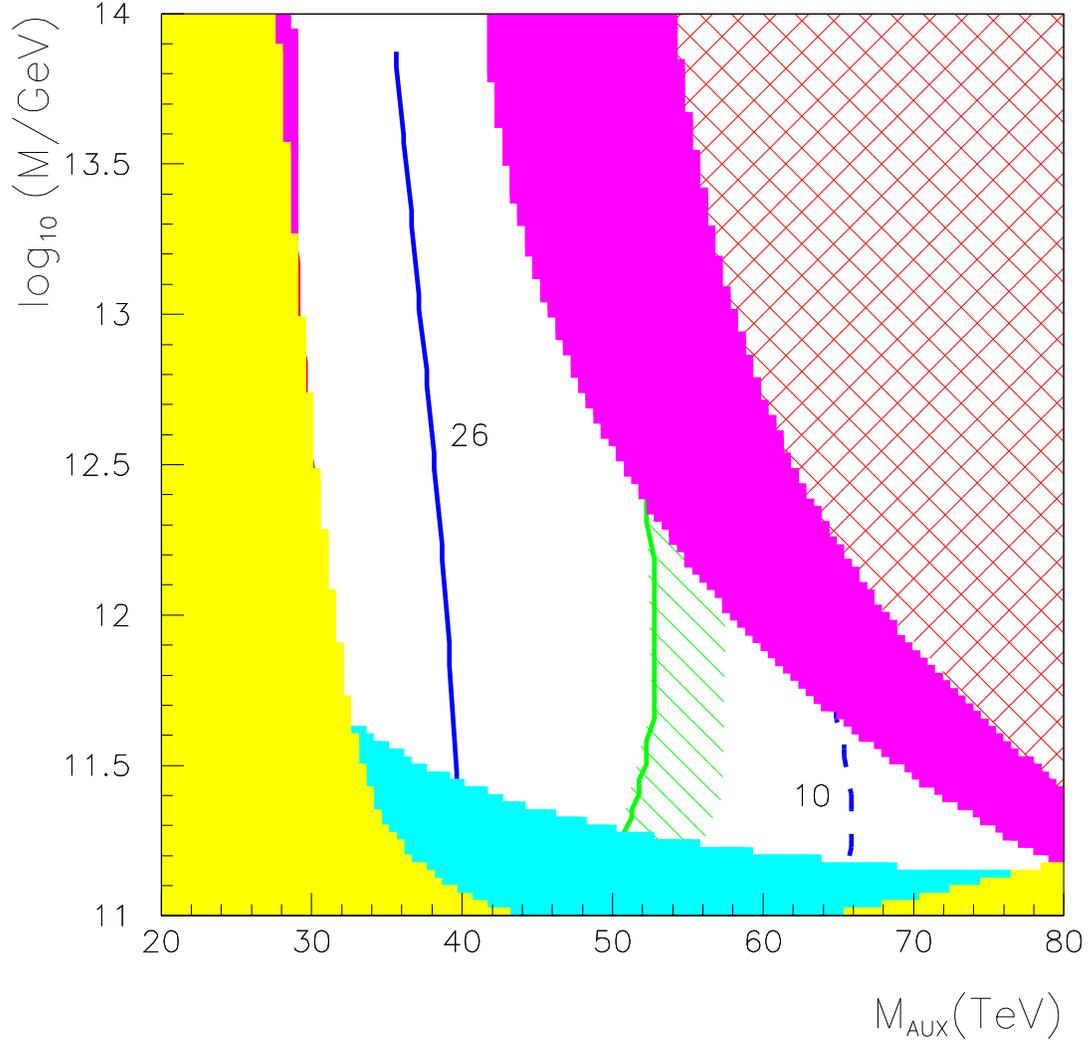, bbllx=60, bblly=60, bburx=480, bbury=466}
\end{center}
\caption{Contour plot on the plane of ($M_{\rm aux},M$) in the deflected
anomaly-mediated model with $N=6,\rho=0$ and $\tan\beta=30$.
Yellow, cyan, and purple regions represent the parameter space
forbidden by the lightest Higgs mass bound, the stau mass bound, and
the chargino mass bound, respectively.
Red cross-hashed area is the region forbidden by electroweak
symmetry breaking condition.
The blue solid and dashed lines stand
for the $-1\sigma$ and $-2\sigma$ values of
$a_\mu^{\rm SUSY}$, respectively.
Green shaded area is the region allowed  by $Br(B\to X_s\gamma)$
and the  green solid line corresponds to the
contour of the $2\sigma$ lower bound of $Br(B\to X_s\gamma)$.
}
\end{figure}


\end{document}